\RequirePackage{rotating}
\RequirePackage{wrapfig}
\documentclass[fleqn,usenatbib]{mnras}
\usepackage{graphicx}	
\usepackage{amsmath}	
\usepackage{amssymb}	
\usepackage[graphicx]{realboxes}
\usepackage[T1]{fontenc}
\usepackage{ae,aecompl}
\usepackage{hyperref}
\usepackage{pdflscape}
\usepackage{newtxtext,newtxmath}
\usepackage{wrapfig}
\usepackage{subfig}

\def\simlt{\mathrel{\rlap{\lower 3pt\hbox{$\sim$}}\raise 2.0pt\hbox{$<$}}}
\def\simgt{\mathrel{\rlap{\lower 3pt\hbox{$\sim$}} \raise 2.0pt\hbox{$>$}}}

\title[Far-infrared radio correlation for lensed galaxies]{The far-infrared/radio correlation for a sample of strongly lensed dusty star-forming galaxies detected by \textit{Herschel}}

\author[M. Giulietti et al.]{
M. Giulietti$^{1,2}$\thanks{E-mail: mgiuliet@sissa.it},
M. Massardi,$^{3,4,1}$
A. Lapi$^{1,3,5,6}$,
M. Bonato$^{3,4}$,
A. F. M. Enia$^{2,7}$,
\newauthor
M. Negrello$^{8}$,
Q. D'Amato$^{2,7}$,
M. Behiri$^{1}$,
and G. De Zotti$^{9}$
\\
$^{1}$SISSA, Via Bonomea 265, I-34136 Trieste, Italy\\
$^{2}$INAF - Osservatorio di Astrofisica e Scienza dello Spazio, Via Gobetti 93/3, I-40129, Bologna, Italy\\
$^{3}$INAF/IRA, Istituto di Radioastronomia, Via Piero Gobetti 101, 40129 Bologna, Italy\\
$^{4}$INAF, Istituto di Radioastronomia - Italian ARC, Via Piero Gobetti 101, I-40129 Bologna, Italy\\
$^{5}$INFN - Sezione di Trieste, via Valerio 2, Trieste 34127, Italy\\
$^{6}$IFPU - Institute for fundamental physics of the Universe, Via Beirut 2, 34014 Trieste, Italy\\
$^{7}$University of Bologna - Department of Physics and Astronomy “Augusto Righi” (DIFA), Via Gobetti 93/2, I-40129, Bologna, Italy \\
$^{8}$School of Physics and Astronomy, Cardiff University, The Parade, Cardiff CF24 3AA, UK\\
$^{9}$INAF, Osservatorio Astronomico di Padova, Vicolo Osservatorio 5, I-35122 Padova, Italy\\
}

\date{Accepted XXX. Received YYY; in original form ZZZ}

\pubyear{2021}

\begin{document}
\label{firstpage}
\pagerange{\pageref{firstpage}--\pageref{lastpage}}
\maketitle

\begin{abstract}
We investigate the radio-far infrared (FIR) correlation for a sample of $28$ bright high-redshift ($1 \lesssim z \lesssim 4$) star-forming galaxies selected in the FIR from the \textit{Herschel}-ATLAS fields as candidates to be strongly gravitationally lensed. The radio information comes either from high sensitivity  dedicated Australia
Telescope Compact Array observations at $2.1$ GHz or from cross-matches with the FIRST survey at $1.4$ GHz. By taking advantage of source brightness possibly enhanced by lensing magnification, we identify a weak evolution with redshift out to $z\lesssim 4$ of the FIR-to-radio luminosity ratio $q_{\rm FIR}$. We also find that the $q_{\rm FIR}$ parameter as a function of the radio power $L_{1.4\,\rm GHz}$ displays a clear decreasing trend, similarly to
what is observed for optically/radio selected lensed quasars found in literature, yet covering a complementary region in the $q_{\rm FIR}-L_{1.4\,\rm GHz}$ diagram. We interpret such a behavior in the framework of an in-situ galaxy formation scenario, as a result of the transition from an early dust-obscured star-forming phase (mainly pinpointed by our FIR selection) to a late radio-loud quasar phase (preferentially sampled by the optical/radio selection). 
\end{abstract}


\section{Introduction}

For several decades the rest-frame 1.4 GHz radio luminosity density $L_{\text{1.4GHz}}$ (W Hz$^{-1}$) emitted by bright dusty star-forming galaxies has been associated to their far infrared (FIR) luminosity $L_{\text{FIR}}$ (8-1000 $\mu$m rest-frame) in terms of an empirical relation between these two quantities, namely the \textit{FIR-radio correlation} (\citealt{Helou1985}, \citealt{Yun2001}). Such a relation is found to be roughly linear across $\sim 3$ orders of magnitude in luminosity $9\lesssim \log{(L_{\text{FIR}}/L_{\odot})} \lesssim 12.5$ with a rather low $1\sigma$ scatter $\lesssim 0.26$ dex; it is often described via the parameter $q_{\text{FIR}}$ (e.g. \citealt{Yun2001}, \citealt{Magnelli2015}) defined as:
\begin{equation}\label{eq:q_ir}
q_{\rm FIR}=\log\left(\frac{L_{\rm FIR}[\hbox{W}]/3.75\times 10^{12}}{L_{1.4\,\rm GHz} [\hbox{W}\,\hbox{Hz}^{-1}]}\right).
\end{equation}
The tightness of this relation can be ascribed to the common origin of the radio and FIR emissions as interpreted by \textit{calorimetric} models (\citealt{Lacki2009}). In this framework, galaxies are assumed to be optically-thick to the UV light coming from young new born stars, which is absorbed by the dust in the interstellar medium and then re-radiated in the FIR regime. At the end of their lives, the same massive stars explode as Type II supernovae producing cosmic ray electrons and positrons, radiating most of their energy in the radio band through synchrotron emission before escaping the galaxy. Additionally, a secondary component of the radio emission comes from the free-free contribution originated by the hot and ionized HII regions.

The FIR-radio correlation is well established in the local universe (\citealt{Helou1985}, \citealt{Yun2001}, \citealt{Jarvis2010}, \citealt{Smith2014}, \citealt{molnar2017}, \citealt{wang2019}). Its apparent tightness encouraged the use of radio emission as an unbiased tracer of obscured star formation in dusty galaxies (\citealt{Kennicutt2012}), and prospectively as a probe to obtain a comprehensive view of the cosmic star formation history up to very high redshift (\citealt{Madau2014}, \citealt{Novak2017}, \citealt{Delhaize2017}). 
This is, in fact, one of the key science drivers of the SKA and of its pathfinder telescopes such as ASKAP and MeerKAT (e.g., \citealt{Jarvis2015}) and of the next-generation Very Large Array
(ngVLA), focused on the investigation of the emission mechanisms that power the radio continuum emission in galaxies (\citealt{Murphy2019}).

For example, early science data at 1.3 GHz from the MeerKAT International GHz Tiered Extragalactic Exploration (MIGHTEE, \citealt{Jarvis2016}) survey have been recently analysed in \cite{An2021}, together with VLA and GMRT radio data, to investigate the radio spectral properties and the FIRRC for a sample of 2094 SFGs in the COSMOS field.
Moreover, future observations with SKA will enable detailed investigations of magnetic fields in galaxies, which can be particularly relevant in the FIRRC especially in the low star formation rate (SFR) regimes (\citealt{Schleicher2016}).

To achieve this goal, however, it is necessary to set on a firm basis the connection between radio and FIR emission (hence star formation), and to assess its redshift dependence, which has been strongly debated in last decades.
From a theoretical perspective, an evolution in redshift is expected as a consequence of either the additional energy losses of the cosmic rays interacting with the cosmic microwave background photons through inverse Compton scattering (e.g \citealt{Murphy2009}; \citealt{Lacki2009}) or because of the co-evolution of AGNs in dusty environments with ongoing star formation in the early stages of galaxy evolution (\citealt{Lapi2018}). In fact, sources whose radio emission is dominated by nuclear activity show up as outliers of the FIR-radio correlation (\citealt{Sopp1991}, \citealt{Stacey2018}). A $q_{\rm FIR} = 1.8$ was proposed by \cite{Condon2002} as the boundary between star-formation and (radio-loud) AGN-dominated radio emission. 

From an observational point of view, the evolution is far from being settled. On the one hand, different works found no significant evidence of a trend with redshift (e.g. \citealt{Sargent2010}). On the other hand, several authors pointed out the presence of a slight decline of the $q_{\text{FIR}}$ parameter:
\cite{Magnelli2015} report evidence of a weak redshift evolution in a mass-selected sample of galaxies, rendered as $q_{\text{FIR}} \propto (1+z)^{-0.12 \pm 0.04}$; a similar result was found by \cite{Basu2015} for a sample of blue cloud galaxies at $z \leq 1.2$; \cite{Tabatabaei2016} studied the radio continuum emission from the KINGFISH sample of nearby galaxies finding that the FIR to 1-10 GHz luminosity ratio could decrease with the star formation rate, suggesting a decrease of the ratio at high redshifts where mostly luminous/star forming galaxies are detected; more recently, \cite{Ocran2020} analysed the radio properties of 1685 star-forming galaxies selected at 610 MHz with the GMRT, inferring an evolution $q_{\rm FIR} \propto 2.86 \pm 0.04 (1+z)^{-0.20 \pm 0.02}$ up to $z \sim 1.8$; \cite{Delhaize2017} found $q_{\rm IR} \propto (1 + z)^{-0.19 \pm 0.01}$ for galaxies selected with the VLA at $3$ GHz; \cite{Calistro2017} obtained $q_{\rm FIR} \propto (1+ z)^{-0.15 \pm 0.03}$ for a sample of star-forming galaxies obtained with Low Frequency Array (LOFAR) at 150 MHz.

This observed evolution would imply that high-z ($z\gtrsim 1$) star forming galaxies present somehow a more pronounced radio emission (or a lower FIR luminosity) compared to their local counterparts. Other studies instead argued on the possibility that the observed trend with the redshift may be a consequence of selection effects (e.g. \citealt{Sargent2010}, \citealt{Bourne2011}, \citealt{Molnar2021}), which can be ascribed to the difference in depth between radio and FIR surveys and to flux-limited samples or to selections biased towards more massive galaxies, as recently reported by different authors (\citealt{Delvecchio2020}, \citealt{Smith2021}, \citealt{Bonato2021}). In fact, investigations at higher redshifts, carried out up to $z\simgt2$ (\citealt{Ivison2010a}, \citealt{Thomson2014}, \citealt{Magnelli2015}) have been limited by the availability of very deep radio data and/or redshift measurements.
In this sense, selection biases can be minimized in homogeneous populations of FIR/submm galaxies (e.g. \citealt{Algera2020a}). In fact, FIR/submm surveys are poorly affected by dust obscuration and feature almost constant flux densities across a wide range of redshifts ($0 \lesssim z\lesssim10$, \citealt{Blain2002}). The resulting strong negative K-correction allowed the detection of such a population predominantly at high redshifts ($z\sim 2-3$) up to $z\sim 6$. 
It has also become clear that the nuclear activity has a crucial impact on the host galaxy and its interstellar medium through the action of energy/momentum feedback (i.e., radio jets), affecting the scatter of the FIR-radio correlation (\citealt{Sopp1991}) and (probably) its redshift evolution. In this sense, the study of dusty star-forming galaxies in the (sub)mm and radio bands is crucial to characterize the interplay between nuclear activity and star formation; high$-z$ dusty galaxies are however compact, with typical intrinsic sizes of a few tenths of an arcsec (\citealt{Pantoni2021}), hence very hard to resolve. 

In recent years, a big step forward has been made thanks to large-area submm surveys which have been used to efficiently select strongly lensed galaxies at high redshift. Lensing enables the observation of regions in the luminosity-redshift space of these sources, that would be otherwise unattainable with current instrumentation or would require an excessive amount of integration time. Indeed, the magnifications of apparent luminosity and angular size by the effect of a foreground lens, offer the unique possibility of studying down to sub-kpc scales the properties of objects otherwise not exceptionally bright, massive, or peculiar, and belonging to the bulk of the galaxy population at the peak of the cosmic star formation history ($z\sim2$, \citealt{Madau2014}). 
In FIR/submm bands, high-z lensed dusty galaxies are particularly bright, while a negligible signal comes from the foreground lens, which is often a massive evolved elliptical at $z\lesssim 1$. Also the obscuration from the foreground lens that limits the investigation of the background galaxy in the optical is negligible in the FIR/submm domain. 

The capabilities of the \textit{Herschel} Space Observatory to select dusty star-forming lensed galaxies have been amply proved. \cite{Wardlow2013} identified 11 lensed galaxies over the 95 deg$^2$ of the \textit{Herschel} Multi-tiered Extragalactic Survey (HerMES; \citealt{Oliver2012}); other 77 candidate lensed galaxies were found by \cite{Nayyeri2016} in the HerMES Large Mode survey (HeLMS; \citealt{Oliver2012}) and in the \textit{Herschel} Stripe 82 Survey (HerS; \citealt{Viero2014}). The \textit{Herschel} Astrophysical Terahertz Large Area Survey (H-ATLAS; \citealt{Eales2010}) is the widest area (600 deg$^2$) extragalactic survey undertaken by \textit{Herschel} and has provided a sample of more than a hundred thousand dusty galaxies at high redshift.
During the Science Demonstration Phase (SDP), covering  a 16 deg$^2$ portion of the sky, \cite{Negrello2010} selected the first sample of 5 strongly lensed galaxies in H-ATLAS.
\cite{Negrello2017} further exploited this survey to extract a catalog of 80 candidate strongly lensed dusty star-forming galaxies brighter than 100 mJy at 500 $\mu$m. 

In this paper we present the radio properties of $28$ \textit{Herschel}-ATLAS candidate strongly-lensed, dusty star-forming galaxies at redshifts $1 \lesssim z\lesssim 4$ extracted from the original sample by \cite{Negrello2017} (see Sec. \ref{sec:sample}). The radio data have been obtained as the result of the cross matches with the FIRST survey at $1.4$ GHz and of dedicated follow-up at $2.1$ GHz with the Australia Telescope Compact Array (ATCA). We then derive the FIR-radio correlation, its redshift evolution and its luminosity dependence (see Sec. \ref{sec:FIRRC}). We compare our result to that of the lensed quasar sample by \cite{Stacey2018}, highlighting the complementarity of the two selections and presenting a physically-motivated interpretation in the framework of an in-situ galaxy evolution model (see \ref{sec:evolutionary_scenario}). Finally, we summarise our findings (see Sec. \ref{sec:conclusions}).

In this work we adopt the standard flat $\Lambda$CDM cosmology \citep{Plank2020} with rounded parameter values: matter density $\Omega_M = 0.32$, dark energy density $\Omega_{\Lambda} = 0.63$, baryon density $\Omega_b = 0.05$, Hubble constant $H_0=100h $ kms$^{-1}$Mpc$^{-1}$ with $h = 0.67$, and mass variance $\sigma_8 = 0.81$ on a scale of 8 $h^{-1}$ Mpc.

\section{The sample}\label{sec:sample}
\subsection{The H-ATLAS lensed galaxies}\label{subsec:h_atlas_sample}

\cite{Negrello2017} extracted a sample of 80 bright FIR/submm-selected galaxies with flux density $S_{500\mu\rm m}>100\,$mJy from the H-ATLAS fields.
The SPIRE photometry is  obtained from the point-source Herschel/SPIRE catalogues of the H-ATLAS Data Release 1 and 2, described in \cite{Valiante2016} and  \cite{Maddox2017a}. The catalogues have been created identifying the 2.5 $\sigma$ peaks in the SPIRE 250 $\mu$m maps, which are then used as position priors to measure flux densities in the other SPIRE bands. 
The instrumental and confusion noise have been minimized through the use of a matching filter for the creation of the maps from which fluxes have been extracted. Only sources with a signal-to-noise ratio $>4$ in at least one of the three SPIRE bands have been included in the final catalogue.
As described in \cite{Negrello2017}, extended sources have been detected through optical images and treated as contaminants, therefore are not included in the sample of 80 candidate lensed galaxies.
The sample galaxies feature FIR (8-1000 $\mu$m) luminosities in the range $13 \lesssim \log(L_{\rm FIR}/L_{\odot}) \lesssim 14$ (uncorrected for lens magnification). They span a redshift range $1 \lesssim z \lesssim 4.5$ (29 spectroscopic, 51 photometric redshifts) with a median value $z_{\rm med}=2.5$. However, to date, only 21 of the 80 candidates in \cite{Negrello2017} sample have been confirmed to be genuine strongly lensed objects through detailed optical/near-infrared (with HST and Keck) or submm (with SMA) images of their structure, together with spectroscopic redshift measurements of the background lensed galaxy and of the foreground lens. For other 8 sources the lensing scenario is strongly supported by the redshift difference of optical/NIR and submm galaxies along the line of sight. The remaining sources are classified as uncertain because they still miss proper follow-up that could confirm their lensed nature.

Some of the sources in the sample have been individually studied in detail exploiting multi-wavelength high-resolution observations. For example, \cite{Enia2018} performed lens modelling and source reconstruction for 13 of the 500 $\mu$m-selected lensed galaxies in H-ATLAS using high-resolution SMA observations at 870 $\mu$m  (\citealt{Bussmann2013a}).
SDP.81 has been the target of observations and modelling in the ALMA long-baseline campaign (\citealt{Partnership2015}, \citealt{Rybak2015}, \citealt{Rybak2015a},  \citealt{Dye2015},\citealt{Swinbank2015} ,\citealt{Tamura2015a}, \citealt{Hatsukade2015a}, \citealt{Hezaveh2016}). These studies led to the reconstruction of the matter distribution of the foreground lens through the detection of low-mass substructures, together with the pixelated surface brightness distribution of dust in the lensed source. Thanks to ALMA CO and CII spectroscopic data, it was also possible to measure with high precision the gas mass distribution and the kinematics of the clumps, revealing a disturbed morphology of the stellar, gas and dust components (see also \citealt{Rybak2020}). SDP.9 and SDP.11 ALMA high-resolution images of the continuum at 1.1 mm and of CO(6-5) (\citealt{Wong2017}) have been combined with X-ray band observations from Chandra to reconstruct the morphology and characterize the nuclear emission (\citealt{Massardi2017}).

\subsection{ATCA follow-up}

The sample in the H-ATLAS South Galactic Pole field comprises 30 sources with $22<RA<0$h and $-36<\delta<-28$. All these are candidate, i.e. not yet confirmed, strongly lensed galaxies. Observations were centered at 2.1 GHz with a 2 GHz bandwidth and were carried out with the Australia Telescope Compact Array (ATCA) on 2017 December 14th and on 2019 July 31th respectively with a 6 km and a 750 m East-West configurations.The corresponding largest angular scales (LAS) are 3.6 and 1.8 arcmin for the first and second configuration respectively. The expected resolution is $\sim 10$ arcsec if the farthest antenna, CA06, is included in the data, as in our case.
In 2017 we got 12 allocated hours that allowed us to perform $\sim7\times2$min cuts on each target. In 2019 we performed $2\times2$min at least on each source for a total of 4 observing hours. The observing strategy allowed enough coverage of the uv domain to recover suitable images of our targets. The weather and system conditions were excellent in both the observing epochs.

Data editing, calibration and imaging is performed using Miriad (\citealt{Sault1995}). 
Flagging of data is fully automatic and exploits the \texttt{PGFLAG} task, and calibration follows common procedures for 16 cm band data-sets. Manual checking confirms that all the major RFI features are removed from the data, flagging more than 30\% of the data, but still recovering a 0.06 mJy beam$^{-1}$ theoretical noise level (1$\sigma$). PKS1934-638 was observed as bandpass and flux calibrator. Two phase calibrators were observed during the 12 hours and 4 hours observations (PKS0008-421 and PKS2255-282) and their solutions are merged to correct the data phases as a function of time. 

The two data-sets are combined during the imaging process with the task \texttt{UVAVER} in order to improve the uv-sampling and the dynamical range of the final image. 
We perform self-calibration during the imaging process in each observed field, exploiting the presence of several bright and point-like sources in the large field of view (22 arcmin FWHM), allowing the improvement of the overall signal-to-noise. 
Flux densities ($S_{\rm image}$) and image noise ($\sigma_{\rm image}$) are then extracted from the Briggs-weighted continuum images with robust factor of 0.5. The average synthesized beam is $\sim 7.4\times3.9$ arcsec. Noise is computed as
\begin{equation}
    \delta S_{\rm image} = \sqrt{(\sigma_{\rm image})^2 + (0.05 \times S_{\rm image})^2}
\end{equation}
to consider the calibration errors. For later use we convert the ATCA flux densities from $2.1$ GHz to $1.4$  GHz assuming a power law spectrum $S_{\nu}=\nu ^{\alpha}$ with average radio spectral index $\alpha=-0.7\pm 0.14$, consistently with \cite{Stacey2018}.

\begin{figure*}

	\includegraphics[width=5.8cm]{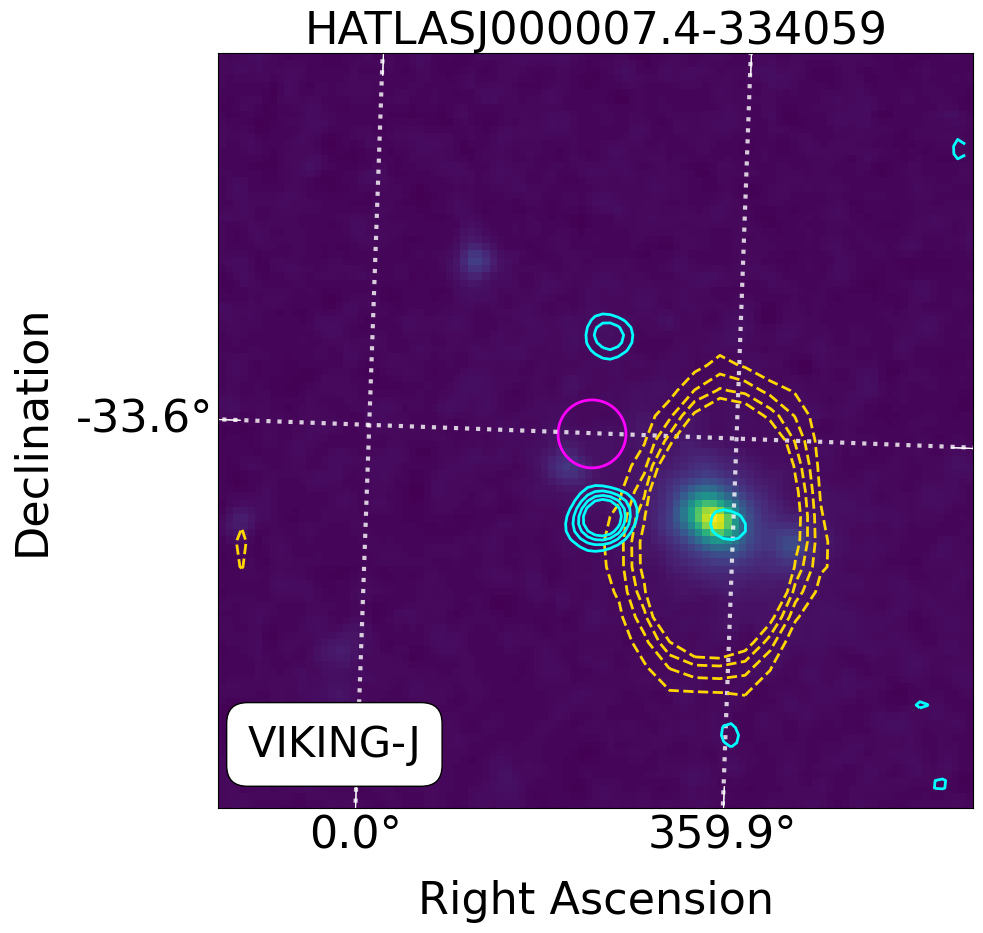}  
	\includegraphics[width=5.8cm]{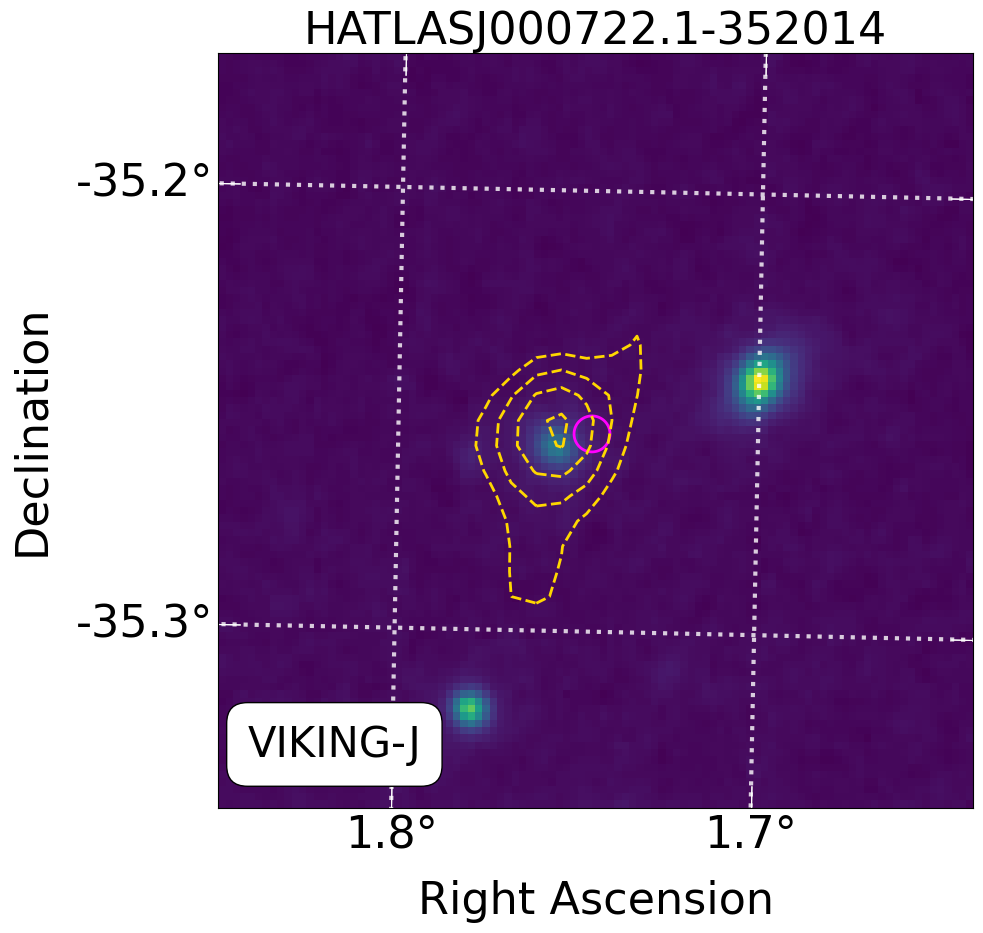} 
	\includegraphics[width=5.9cm]{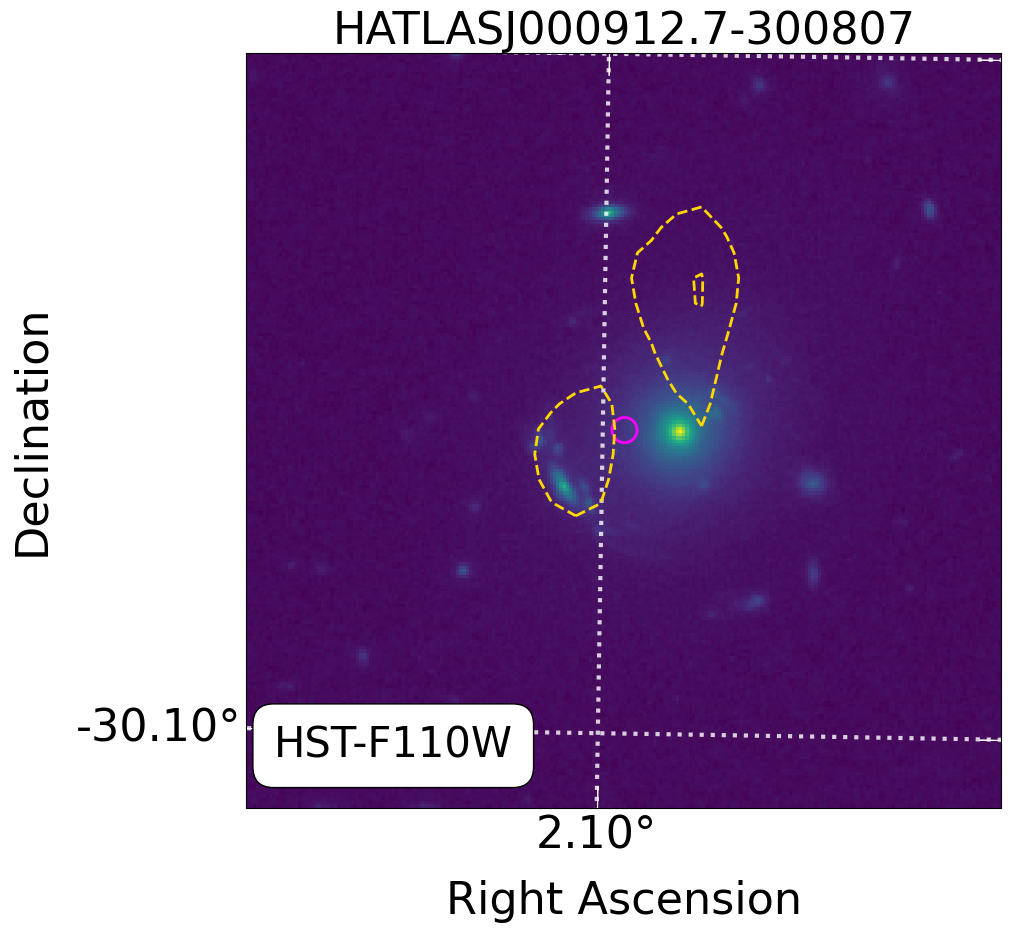} 
	\includegraphics[width=5.95cm]{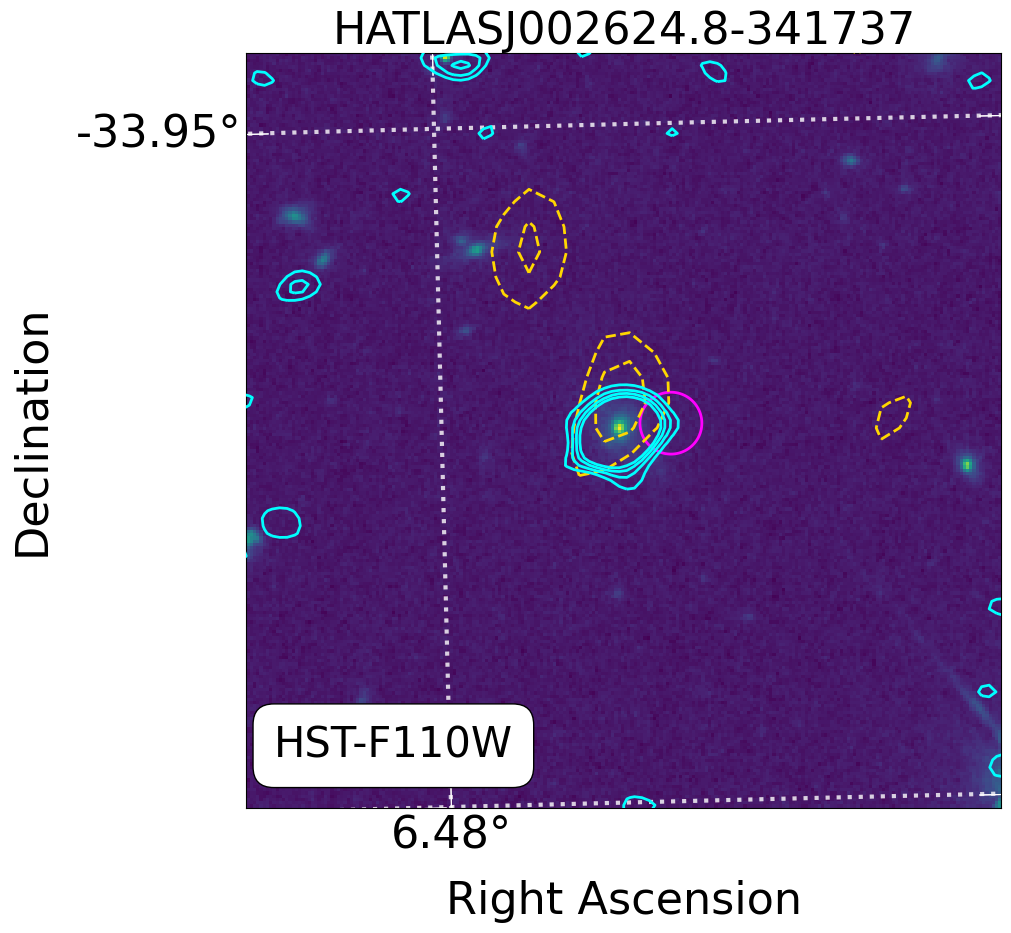} 
	\includegraphics[width=5.8cm]{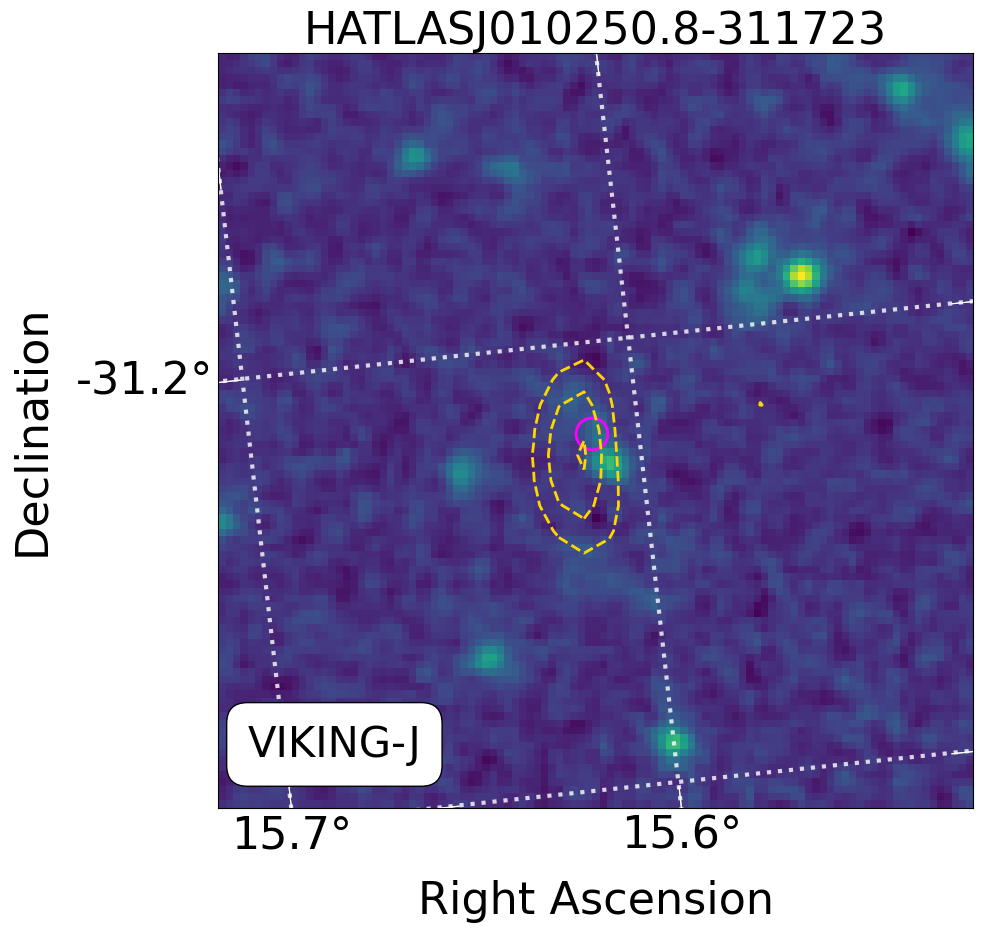} 
	\includegraphics[width=5.8cm]{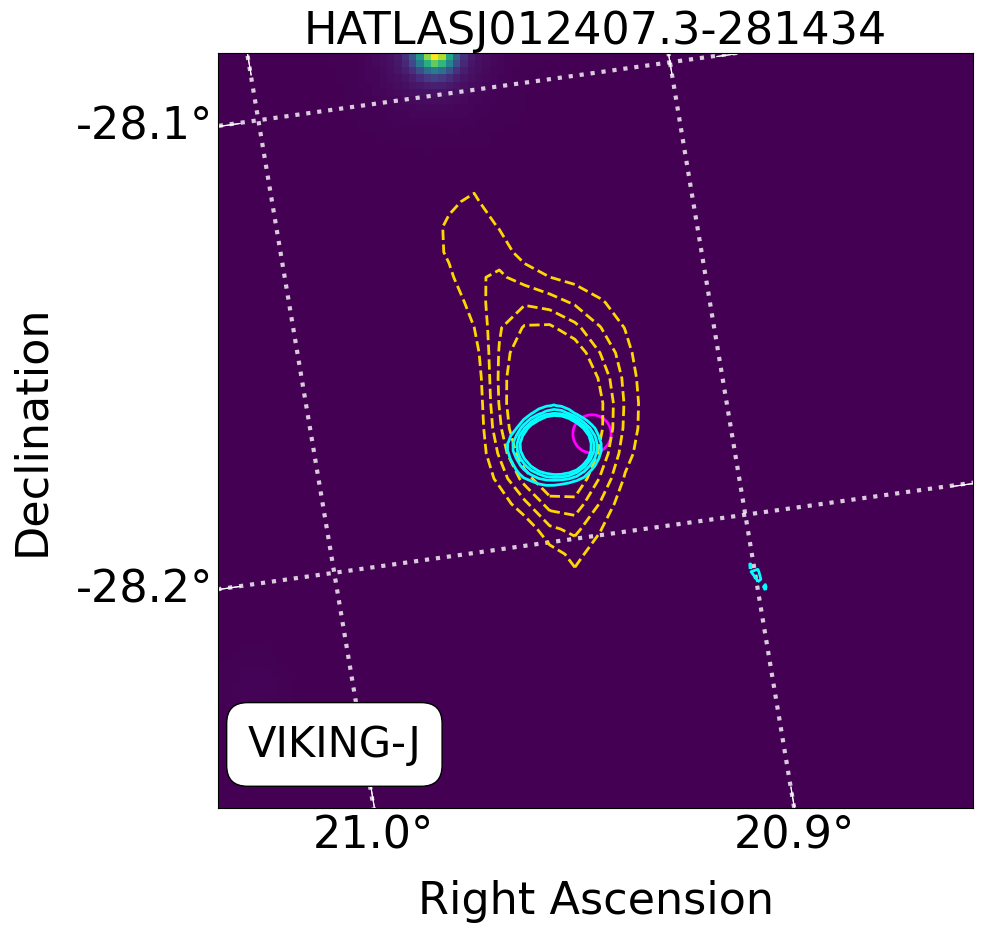}  
	\includegraphics[width=5.8cm]{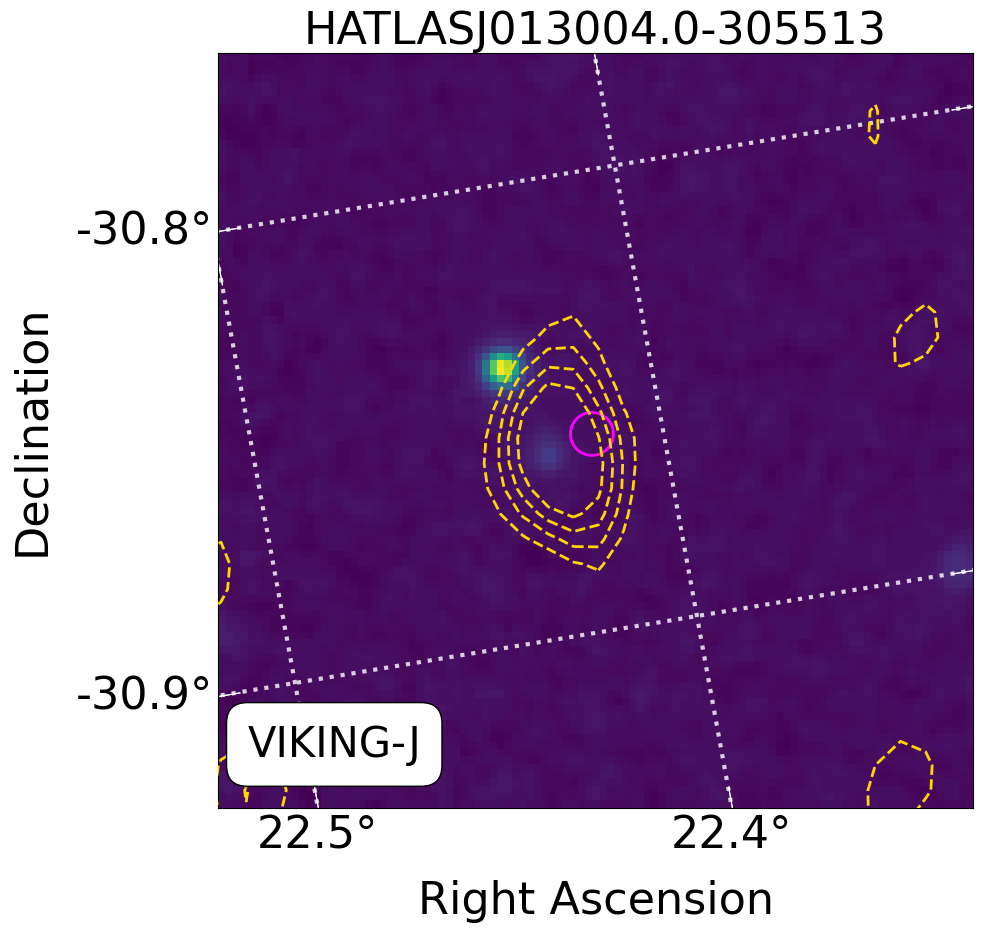}
	\includegraphics[width=5.9cm]{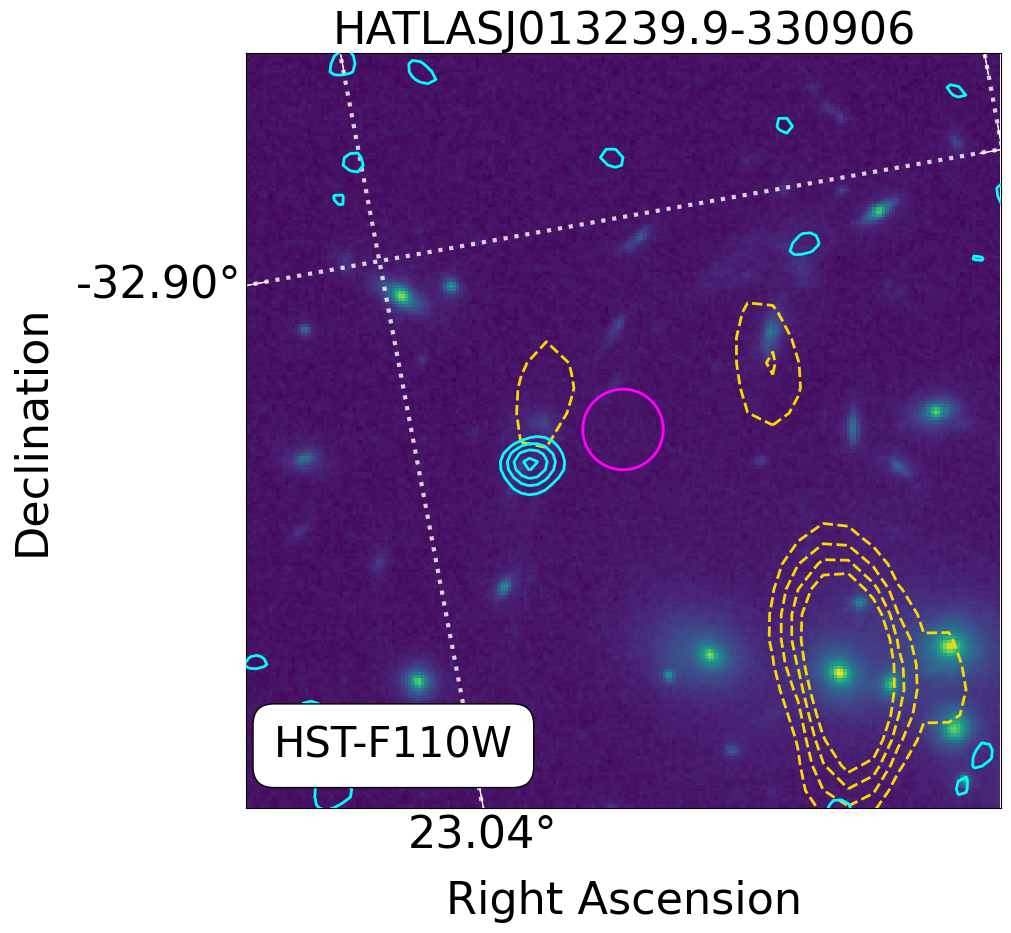}
	\includegraphics[width=5.85cm]{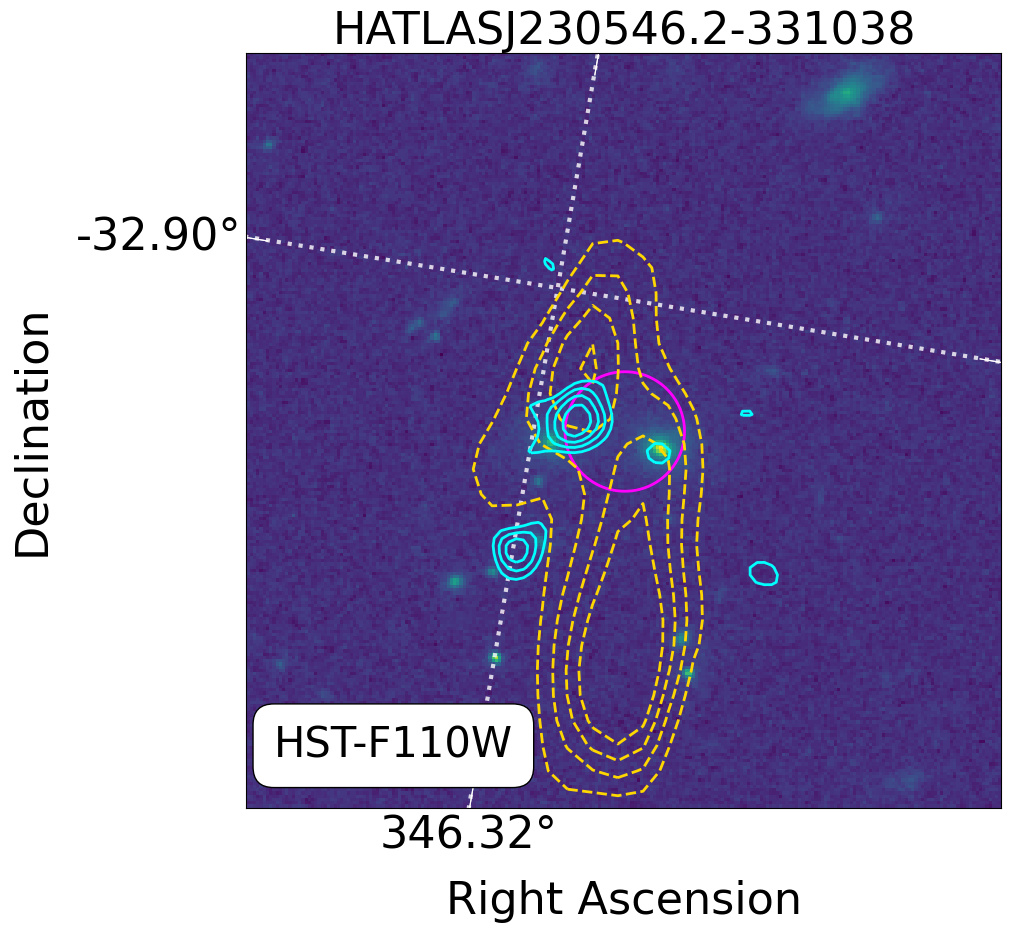} 
	\includegraphics[width=5.9cm]{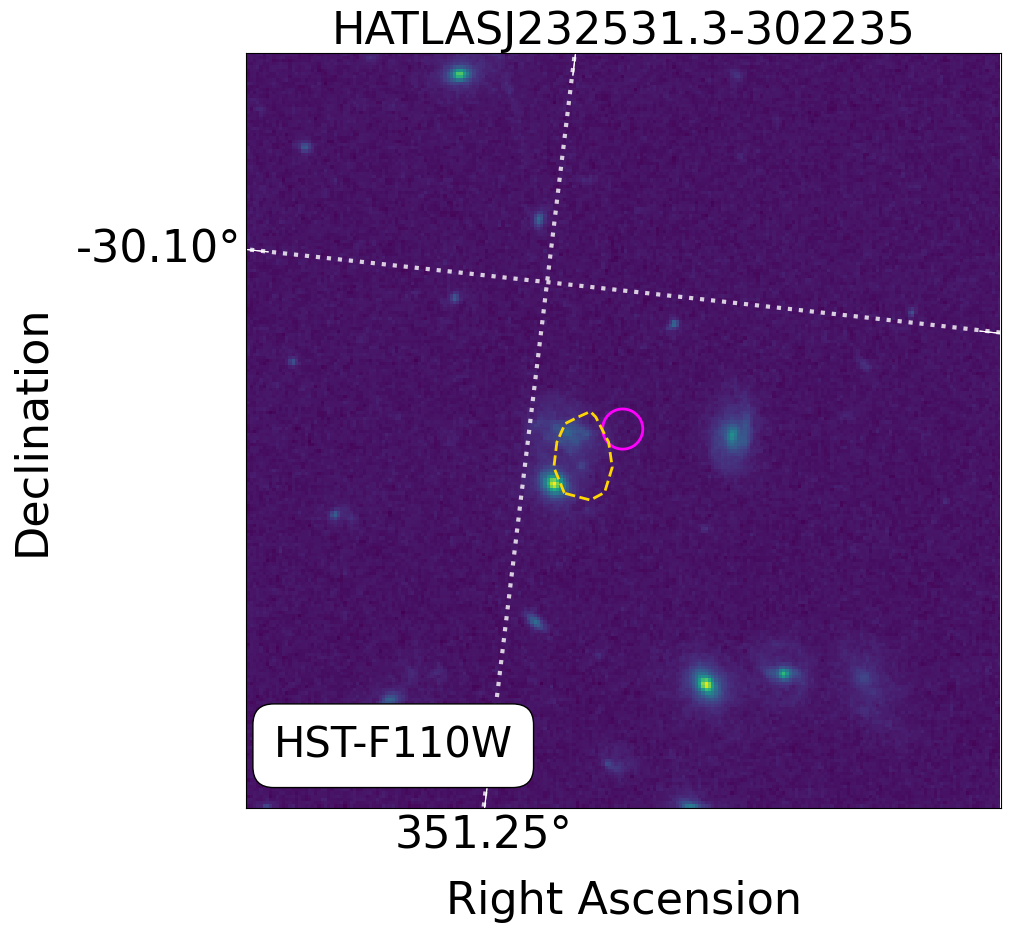} 
	\includegraphics[width=5.8cm]{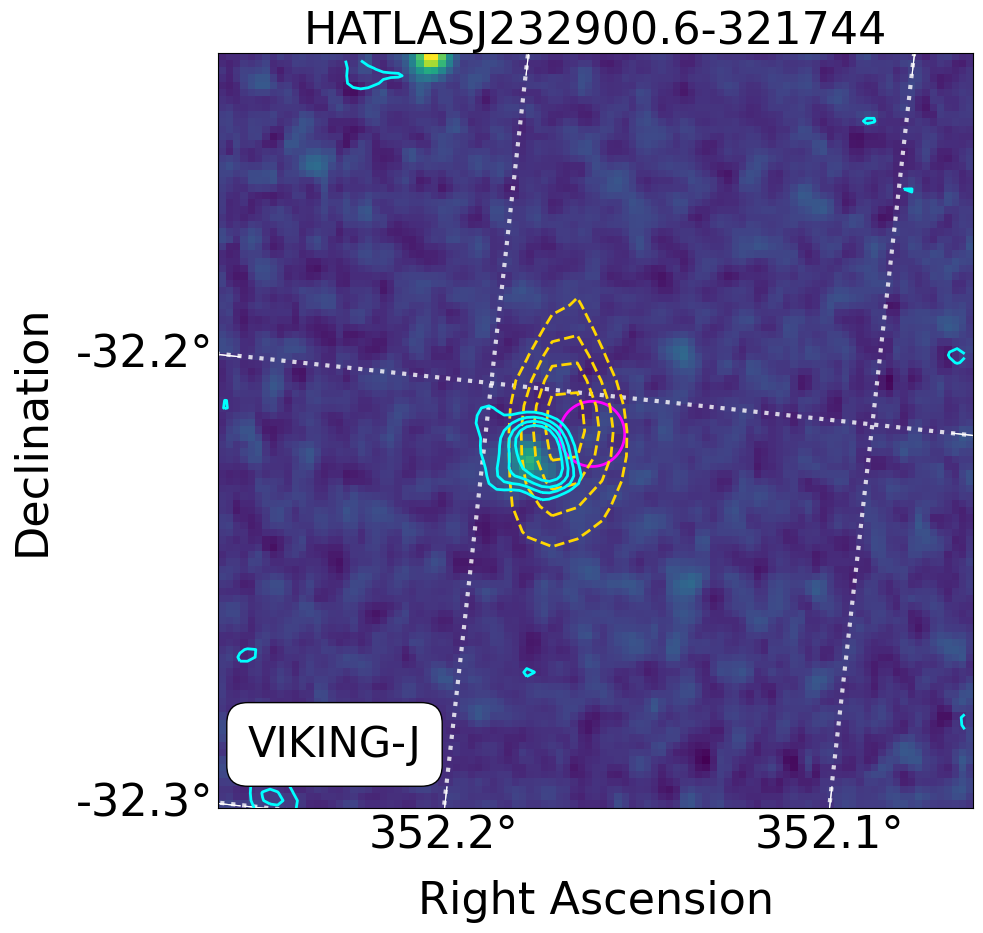} 
    \caption{Cutouts of NIR images with the best available angular resolution centred on the Herschel positions for the 11 sources with an ATCA cross-match. Contours at 3,5,7,9$\sigma$ levels are showed in gold for the radio and cyan for ALMA. Circles are the SPIRE position with a 3$\sigma_{\rm pos}$ radius. RA and Dec are reported in Deg units. The postage stamps are 30x30 arcsec}
    \label{fig:matches}
\end{figure*}

\subsection{FIRST catalogue counterparts}

\begin{figure*}
	\includegraphics[width=5.7cm]{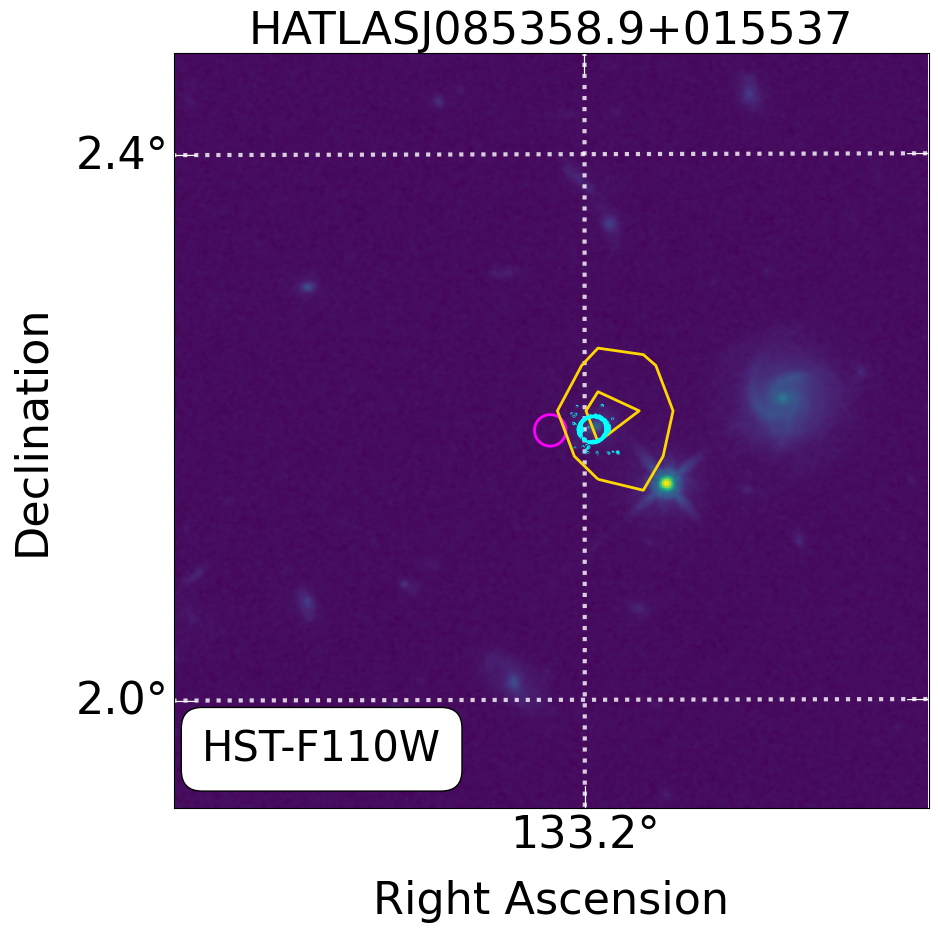}  \includegraphics[width=5.8cm]{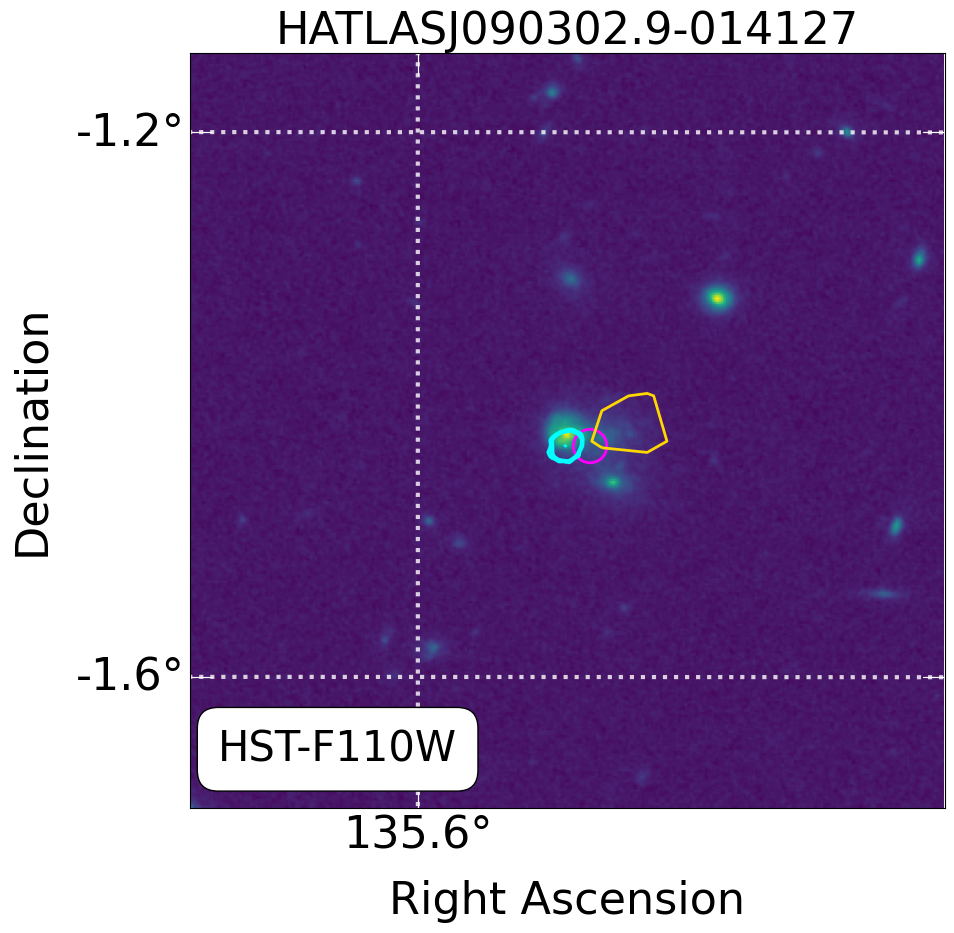} 
	\includegraphics[width=5.9cm]{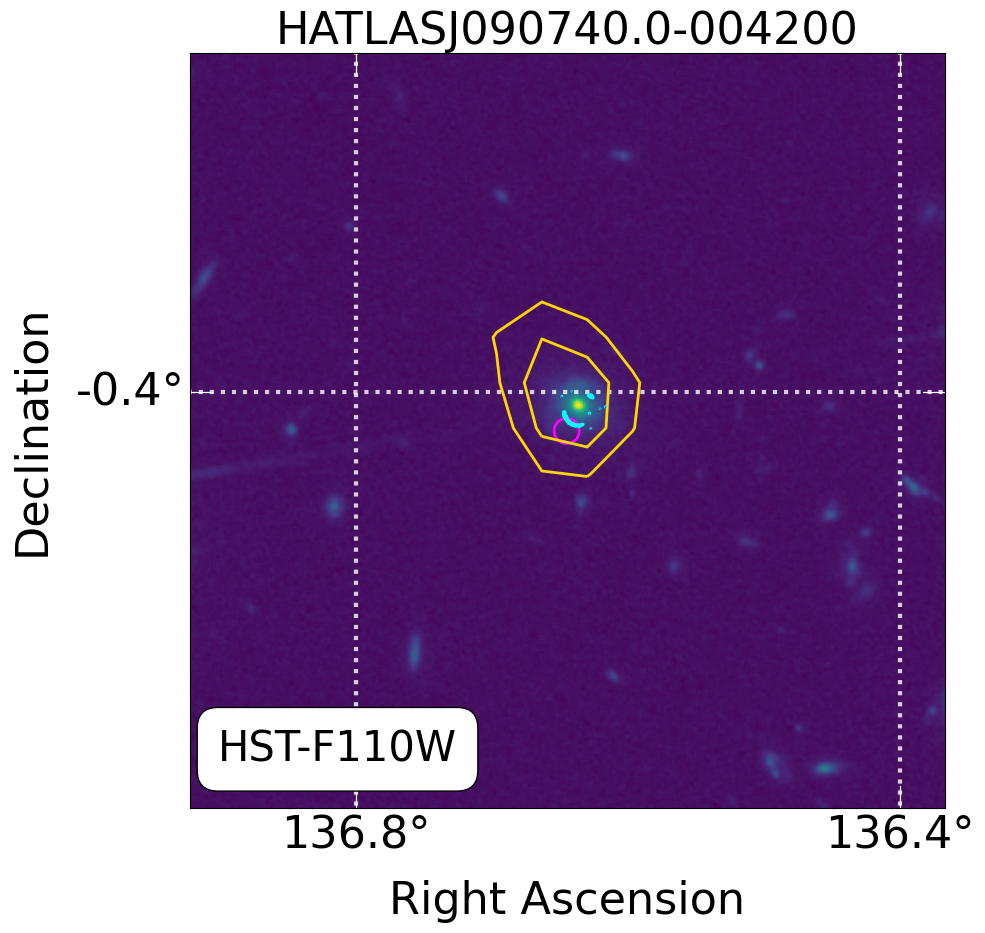} 
    \includegraphics[width=5.7cm]{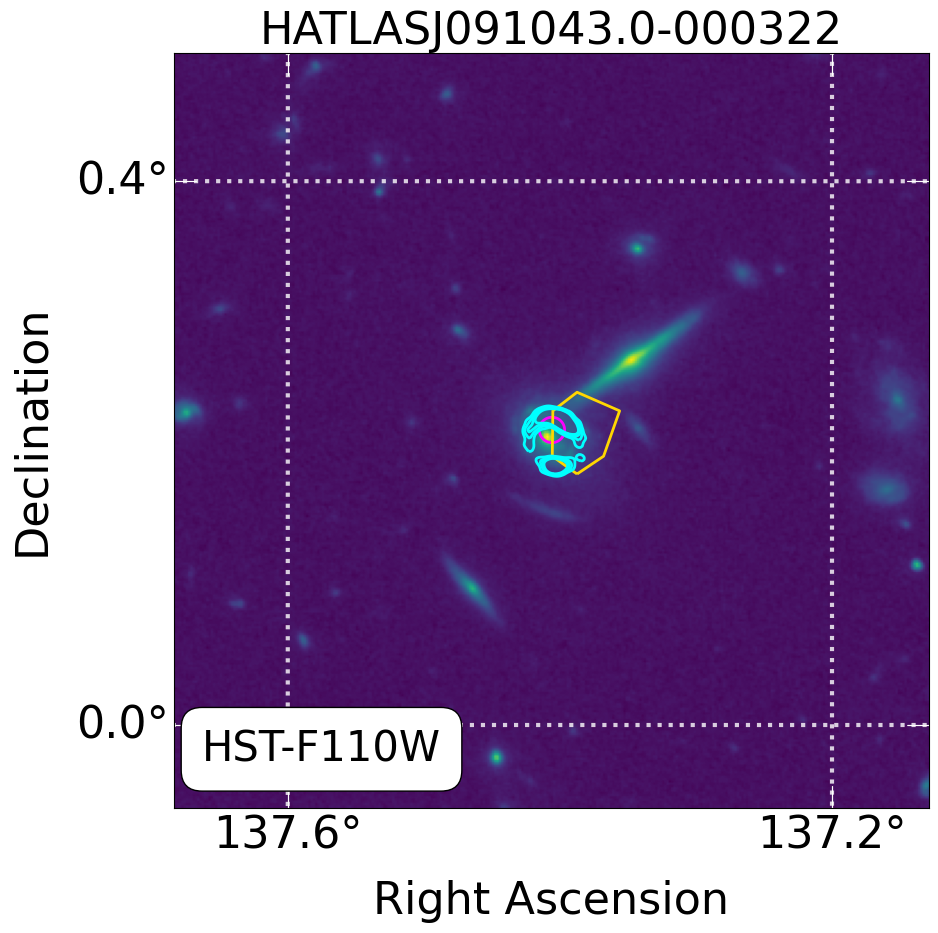} 
	\includegraphics[width=5.7cm]{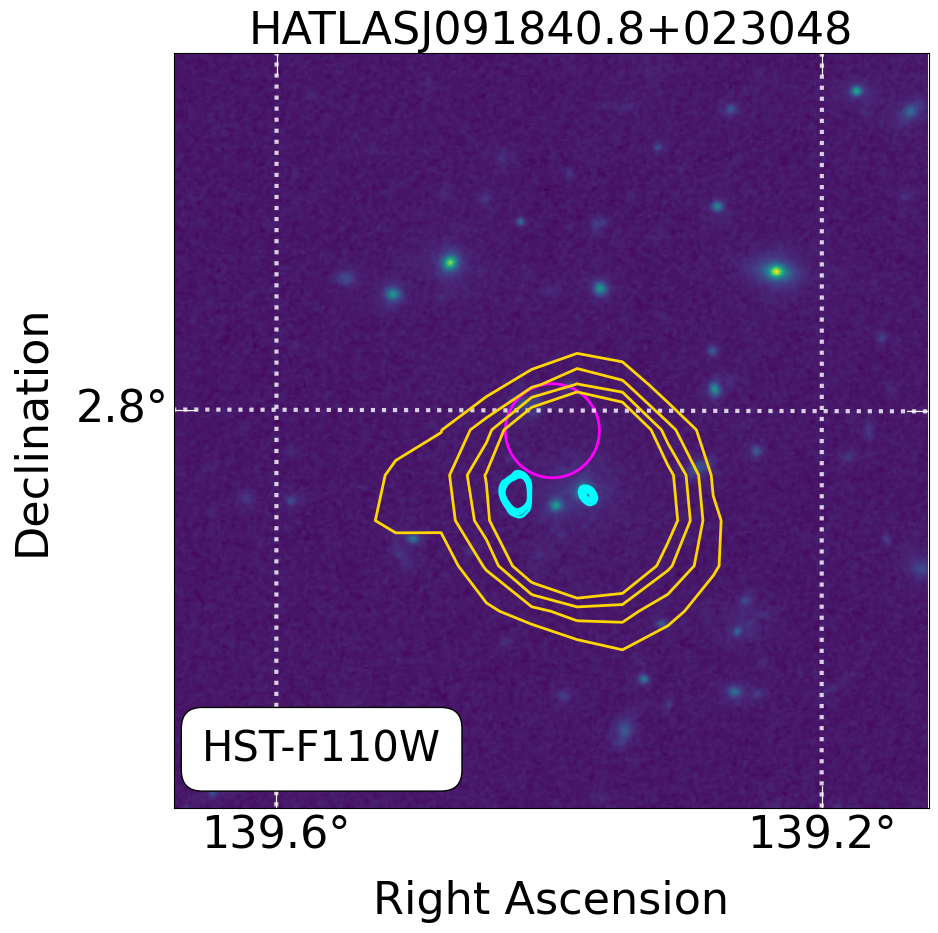} 
	\includegraphics[width=5.7cm]{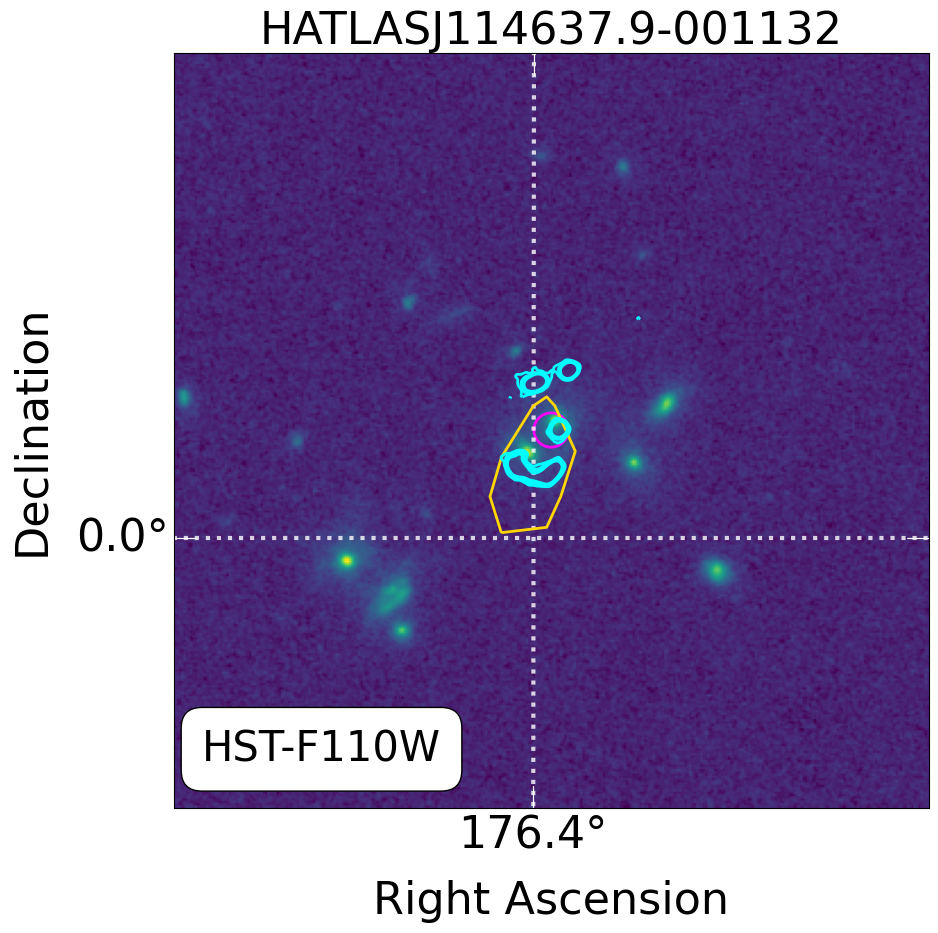} 
	\includegraphics[width=5.7cm]{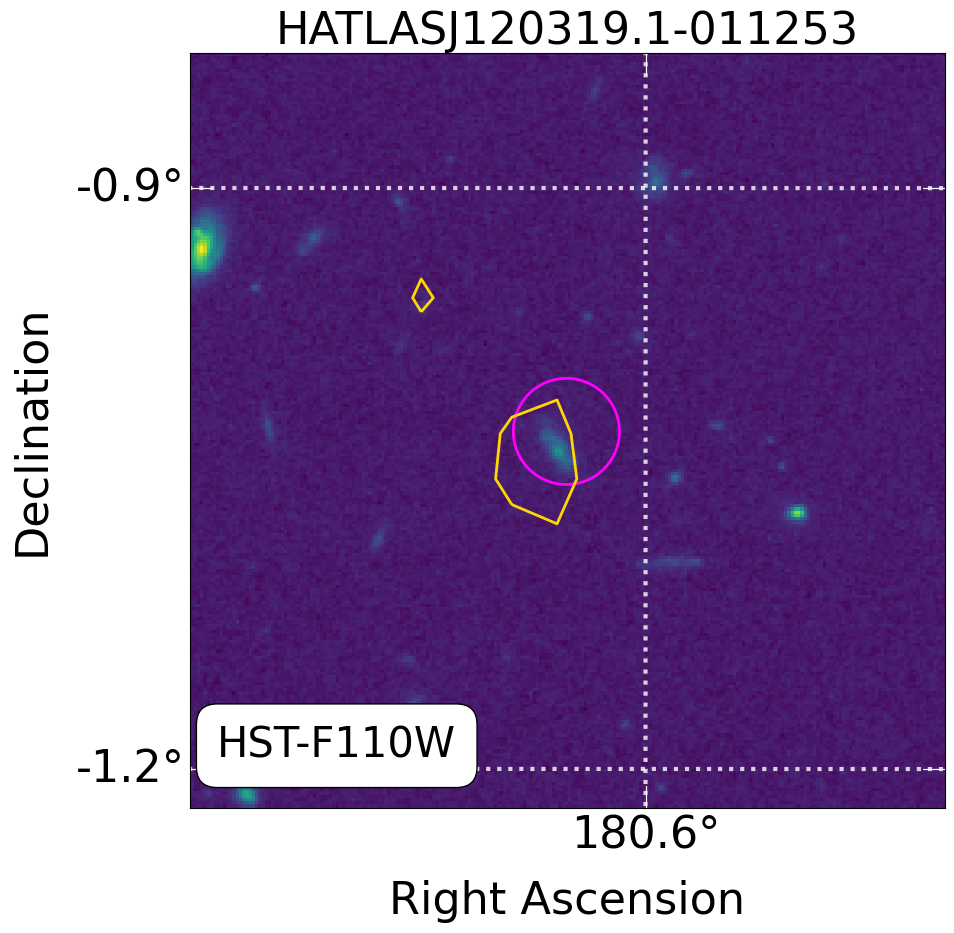}
	\includegraphics[width=5.7cm]{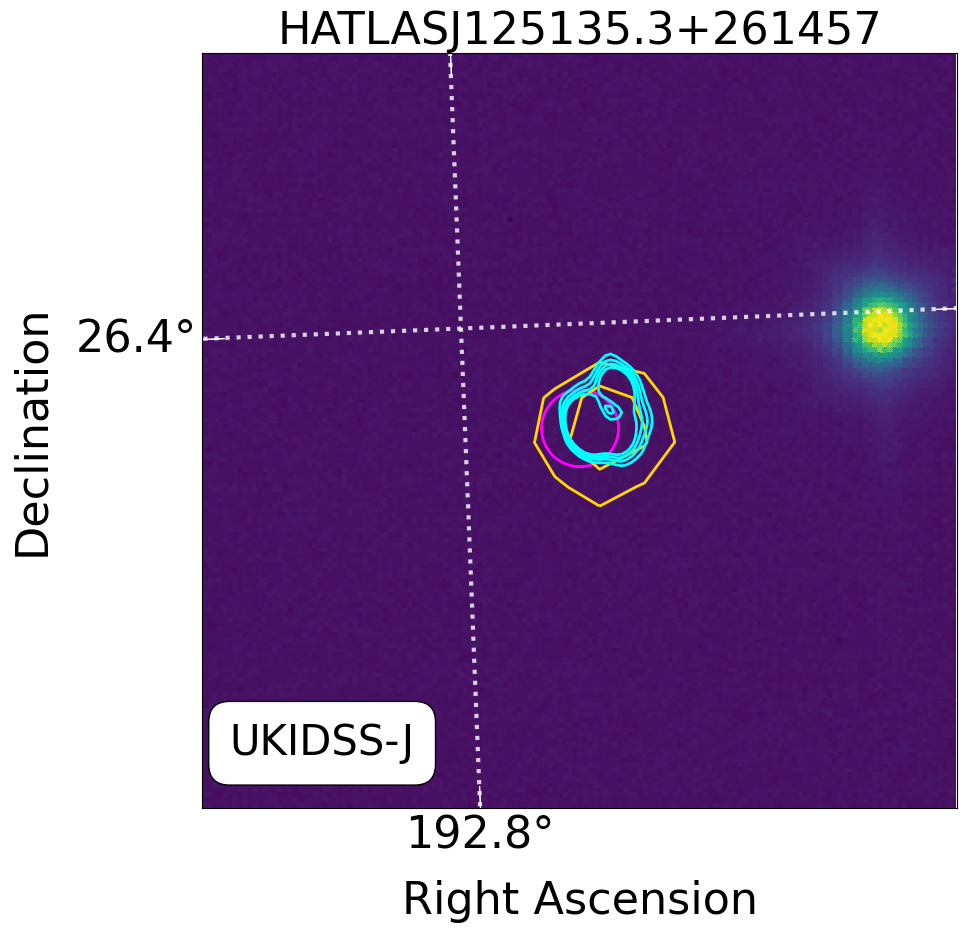}
	\includegraphics[width=5.7cm]{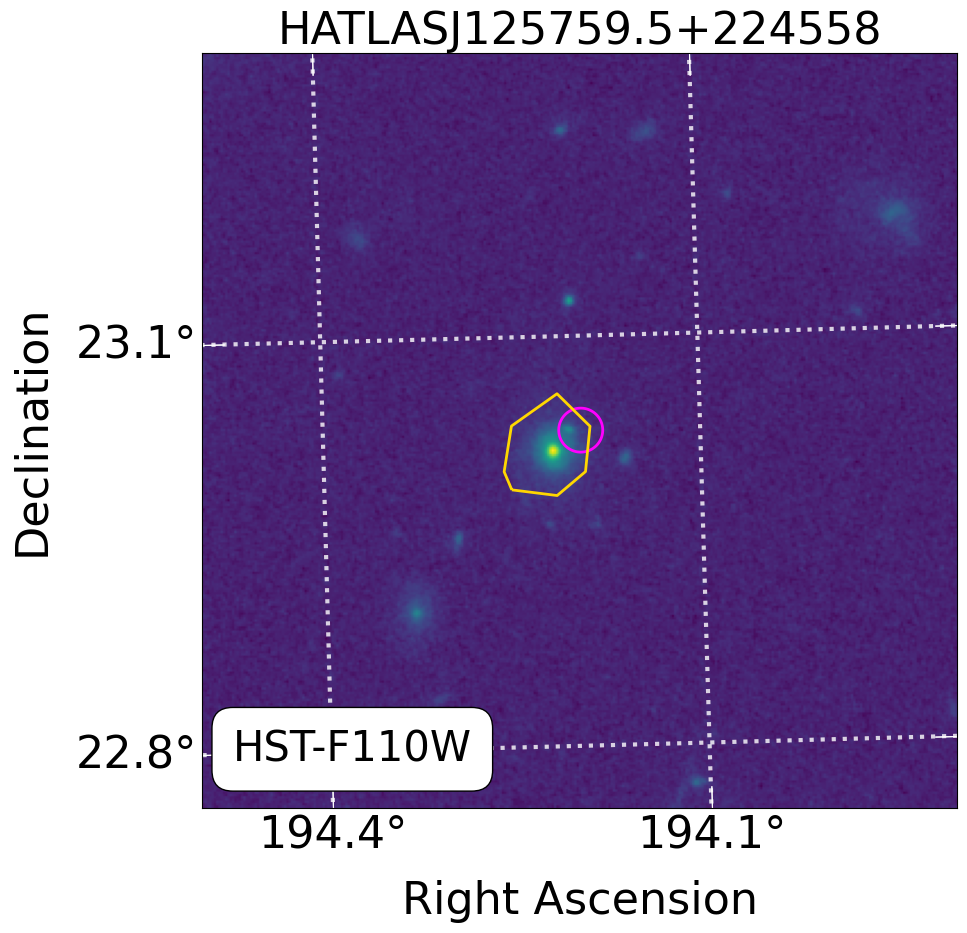}
 	\includegraphics[width=5.7cm]{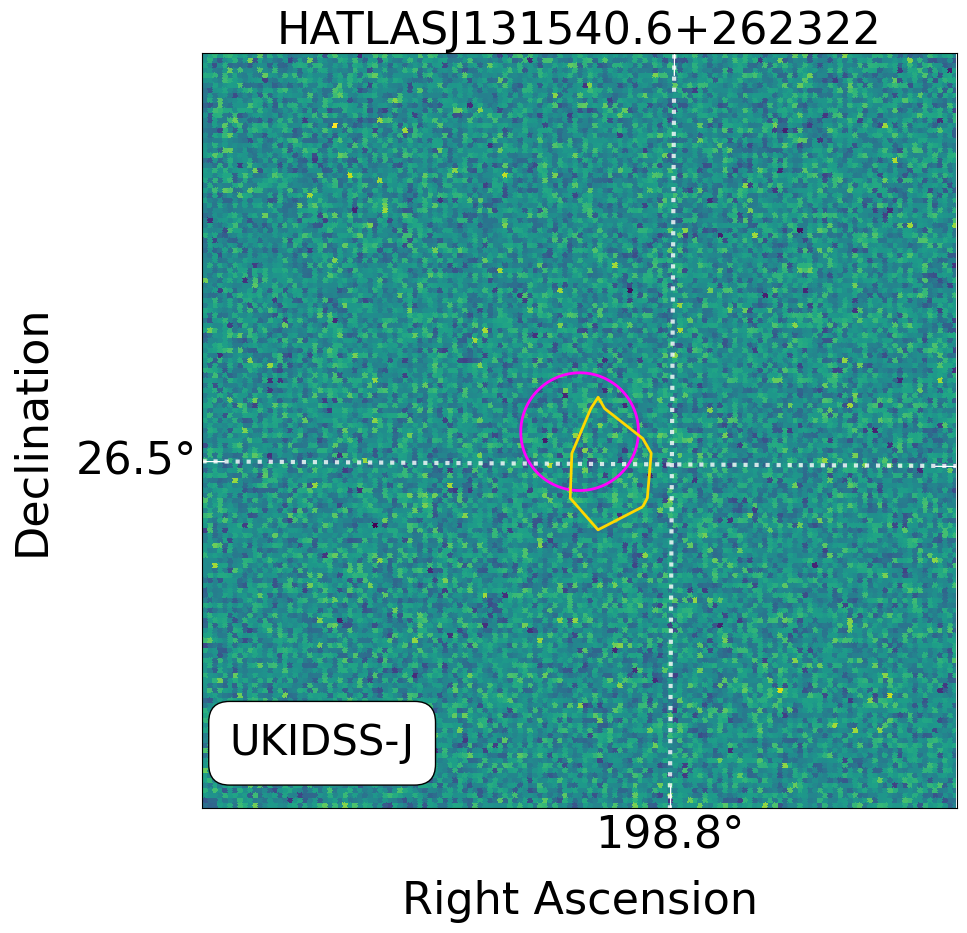}
	\includegraphics[width=5.7cm]{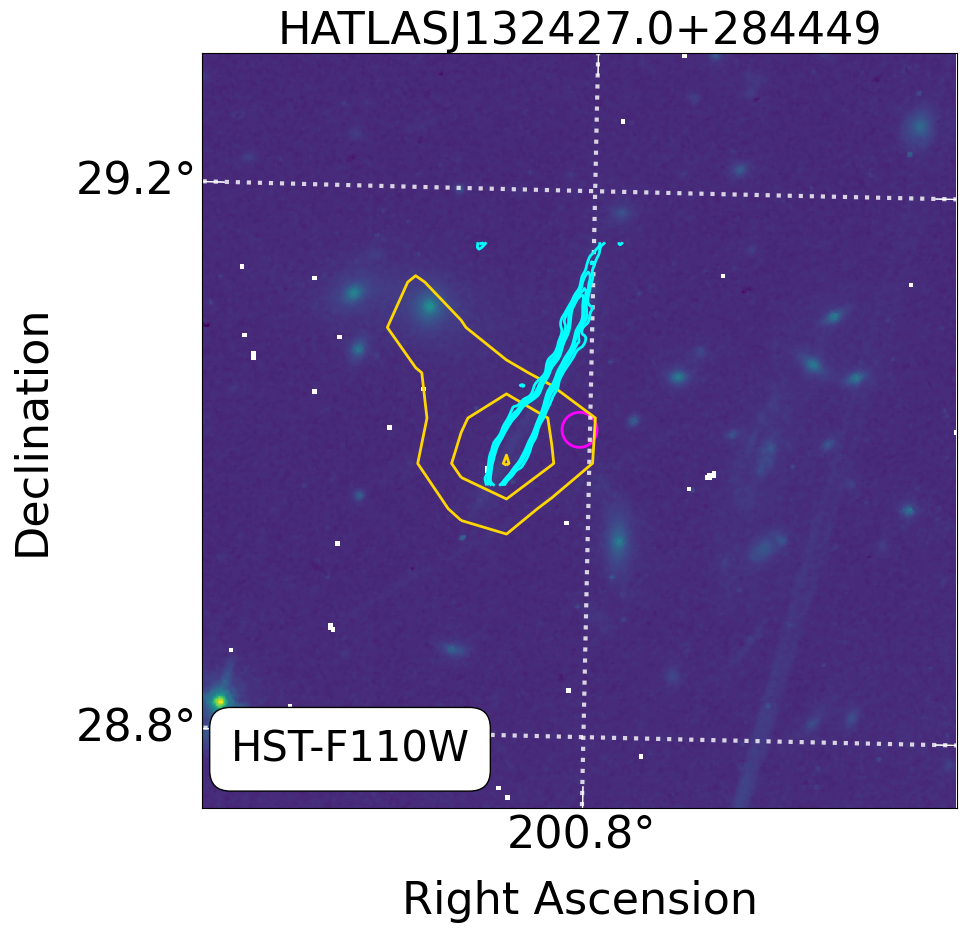}
	\includegraphics[width=5.7cm]{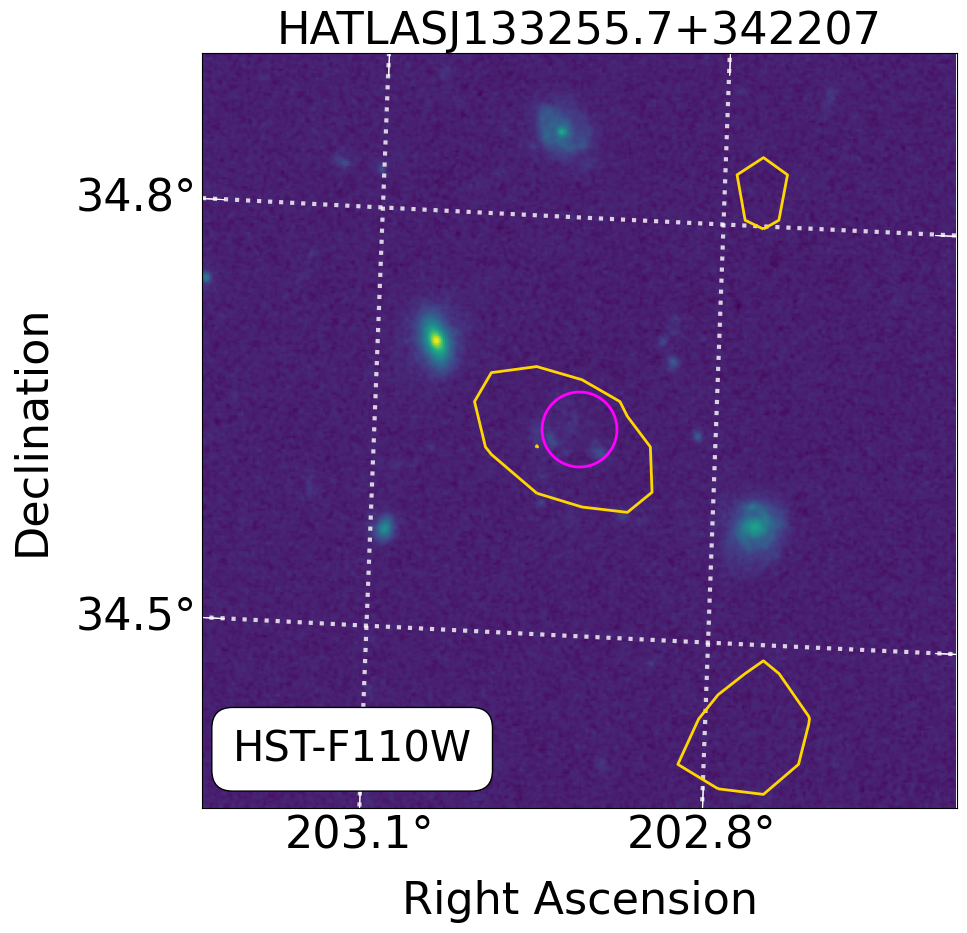}

    \caption{Cutouts of NIR images with the best available angular resolution centred on the Herschel positions for the 16 sources with a FIRST cross-match. Contours at 3,5,7,9$\sigma$ levels are showed in gold for FIRST and cyan for ALMA. Circles are the SPIRE position with a 3$\sigma_{\rm pos}$ radius. RA and Dec are reported in Deg units. The postage stamps are 30x30 arcsec.}
    \label{fig:first_matches}
\end{figure*}

\begin{figure*}
    \ContinuedFloat
	\includegraphics[width=5.7cm]{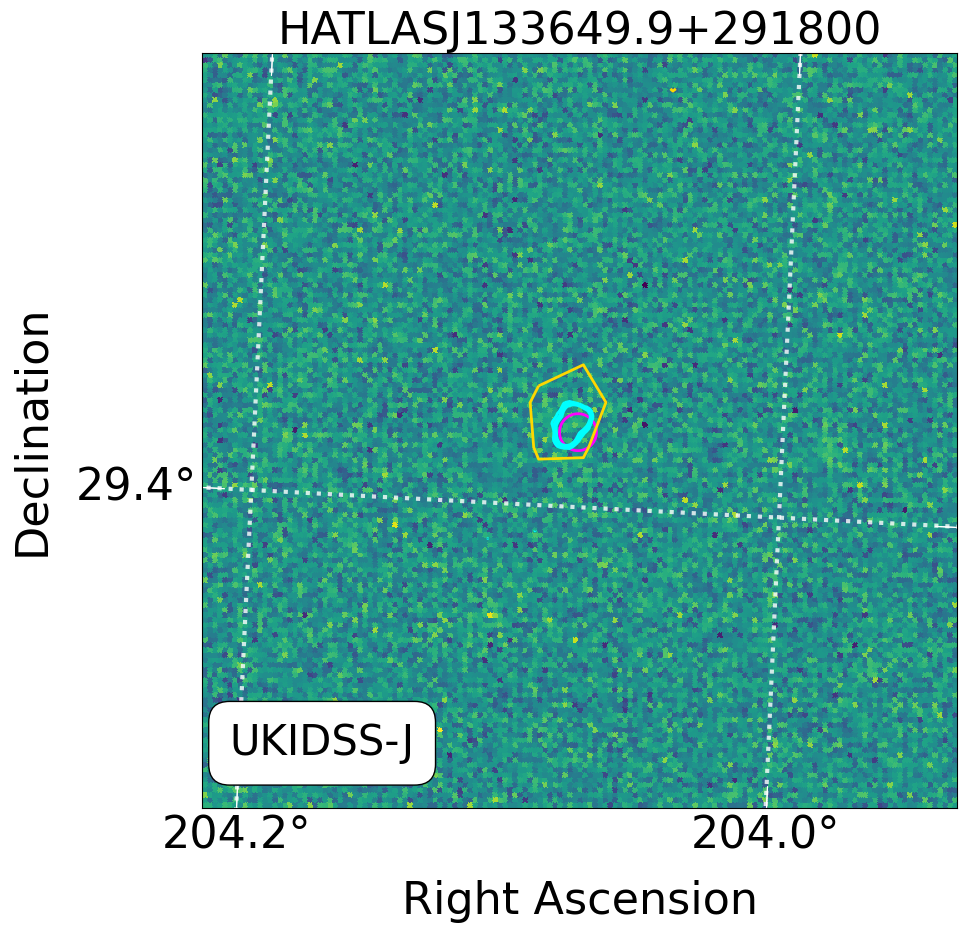}
	\includegraphics[width=5.7cm]{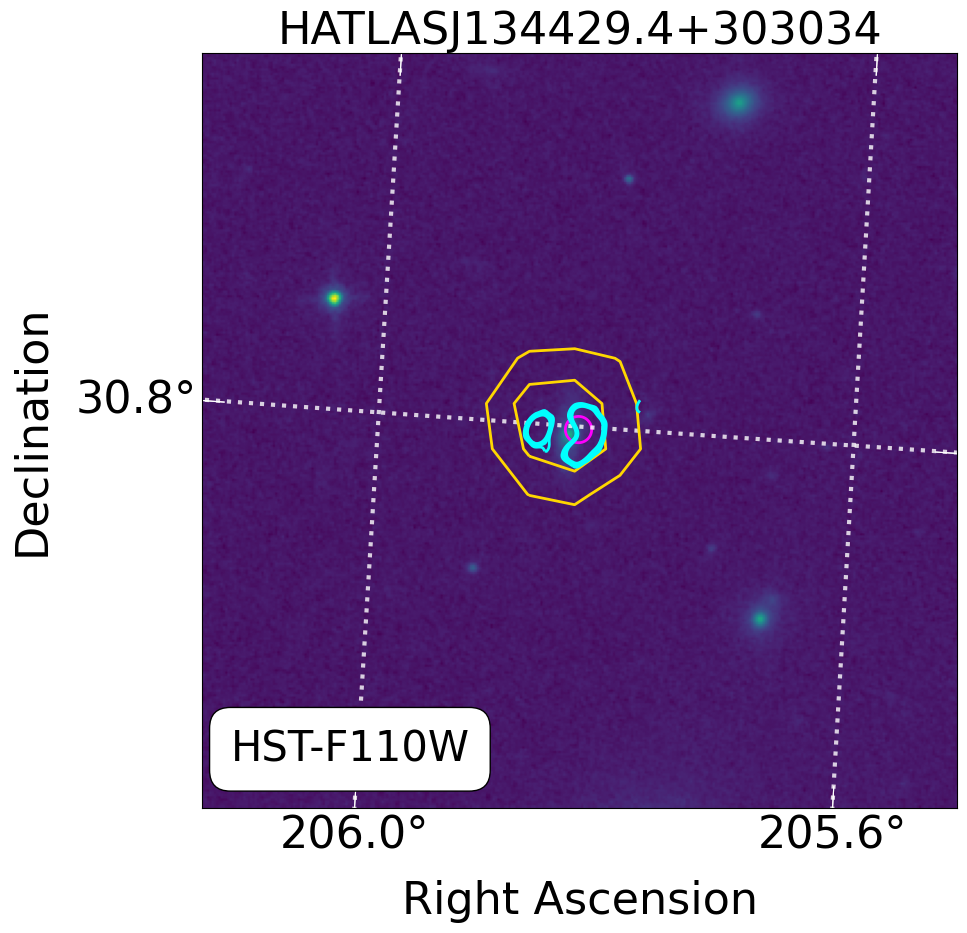}
	\includegraphics[width=5.6cm]{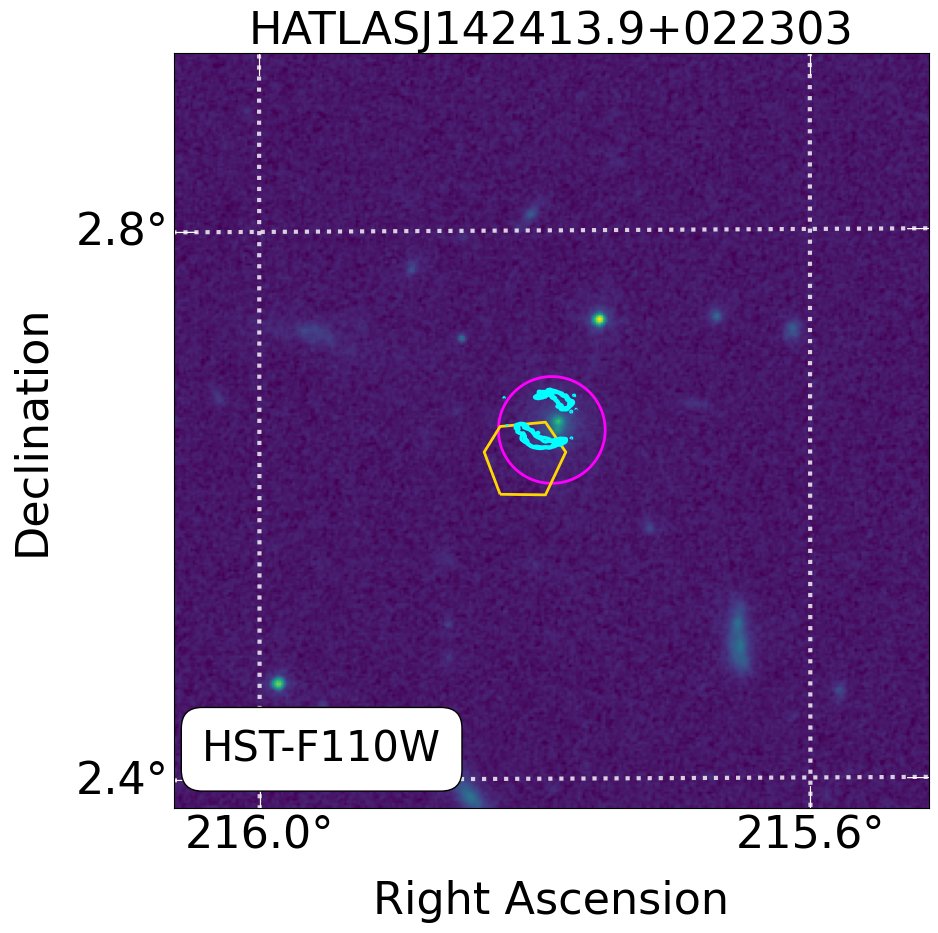}
	\includegraphics[width=5.7cm]{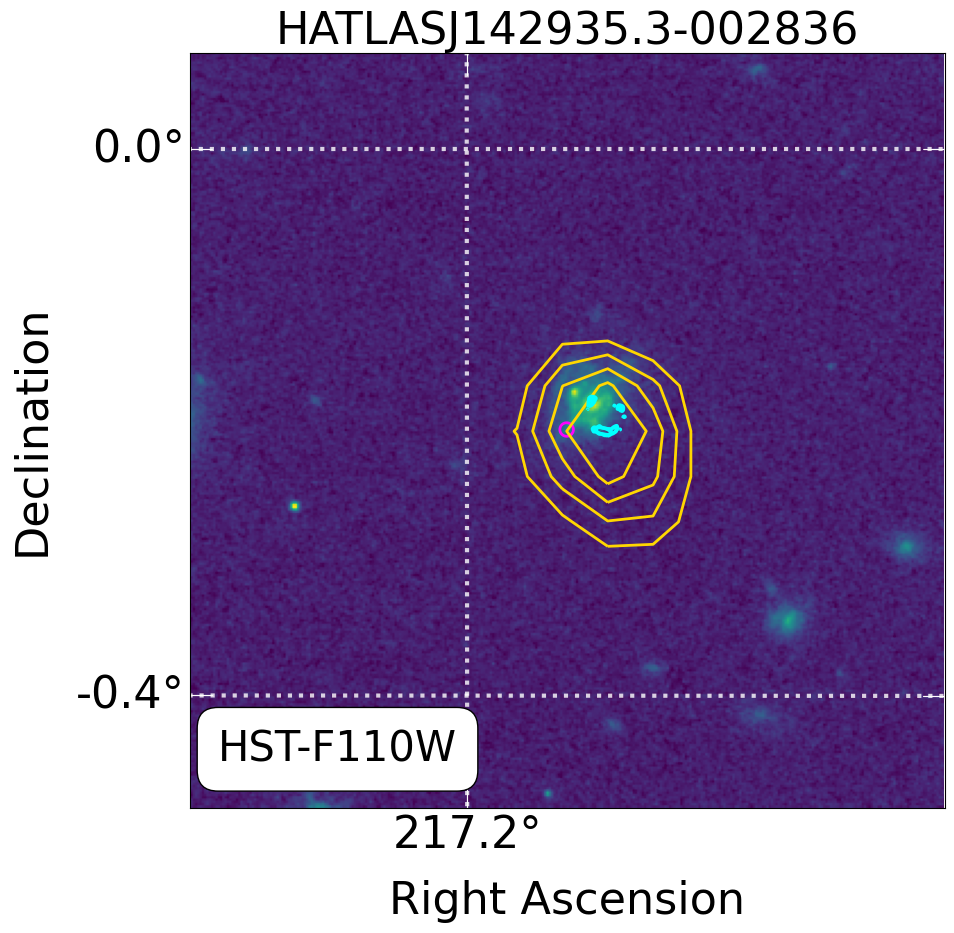}
  \caption{\textit{Continued.}}
\end{figure*}

\begin{table*}
\begin{center}
    
\caption{Radio and FIR properties of the candidate and confirmed lensed galaxies with a counterpart in the radio bands. Columns are as follows: H-ATLAS IDs; \textit{Herschel}/SPIRE flux densities; source redshift; lens classification from \protect\cite{Negrello2017}: (A) confirmed dusty lensed, (B) likely to be lensed,  (C) uncertain; radio flux densities at 1.4 GHz; origin of the radio counterpart (images, catalogue or literature); distance (in arcseconds) of the radio peak from the \textit{Herschel} position, if available.}
\label{tab:matches}
\begin{tabular}{lcccccccc}
\hline
$\#$H-ATLAS ID  &$S_{250\mu m}$ &$S_{350\mu m}$ &$S_{500\mu m}$ &z$^{\dag}$&lenscode &$S_{\rm 1.4GHz}$ &counterpart &dist \\
 &[mJy]&[mJy]&[mJy]& & & [mJy]&  &[arcsec]\\
\hline
J000007.4-334059 &130.3$\pm$5.4 &160.1$\pm$5.9 &116.2$\pm$6.5 &2.56$\pm$0.43 &C & 1.91$\pm$0.2 &ATCA image & 6.6	 \\
J000722.1-352014 &237.3$\pm$5.2 &192.8$\pm$5.6 &107.5$\pm$6.6 &1.46$\pm$0.3 &C &0.39$\pm$0.08 &ATCA image&	1.3	 \\
J000912.7-300807 &352.8$\pm$5.4 &272.6$\pm$6.1 &156.1$\pm$6.8 &1.39$\pm$0.29 &C &0.20$\pm$0.05 &ATCA image& 2.2		 \\
J002624.8-341737 &137.7$\pm$5.2 &185.9$\pm$5.8 &148.8$\pm$6.8 &2.96$\pm$0.48 &C &0.26$\pm$0.06 &ATCA image& 1.8	 \\
J010250.8-311723 &267.9$\pm$5.2 &253.1$\pm$5.7 &168.1$\pm$7.1 &1.92$\pm$0.35 &C &0.29$\pm$0.06 &ATCA image&	0.9	 \\
J012407.3-281434 &257.5$\pm$6 &271.1$\pm$6 &203.9$\pm$6.8 &2.31 $\pm$0.4 &C &0.79$\pm$0.13 &ATCA image& 2.1	 \\
J013004.0-305513 &164.4$\pm$4.3 &147.5$\pm$5.1 &100.6$\pm$5.9 &1.84$\pm$0.34 &C &0.65$\pm$0.11 &ATCA image&	1.7	 \\
J013239.9-330906 &112$\pm$5.5 &148.8$\pm$6.2 &117.7$\pm$7 &2.9 $\pm$0.47 &C &0.19$\pm$0.05 &ATCA image&	3.5	 \\
J085358.9+015537 &396.4$\pm$7.6 &367.9$\pm$8.2 &228.2$\pm$8.9 &2.0925  &A &1.19$\pm$0.16 &FIRST cat &2.0 		 \\
J090302.9-014127(SDP.17) &354.1$\pm$7.2 &338.8$\pm$8.1 &220.2$\pm$8.6 &2.3049  &A &0.41$\pm$0.14 &FIRST image & 2.3		 \\
J090740.0-004200(SDP.9) &477.6$\pm$7.3 &327.9$\pm$8.2 &170.6$\pm$8.5 &1.577  &A &1.21$\pm$0.15 & FIRST cat &2.1 	 		 \\
J091043.0-000322(SDP.11) &420.8$\pm$6.5 &370.5$\pm$7.4 &221.4$\pm$7.8 &1.786  &A &0.97$\pm$0.19 &FIRST image & 1.3		 \\
J091840.8+023048 &125.7$\pm$7.2 &150.7$\pm$8.3 &128.4$\pm$8.7 &2.581  &C &4.63$\pm$0.3 &FIRST cat &2.2 			 \\
J114637.9-001132 &316$\pm$6.6 &357.9$\pm$7.4 &291.8$\pm$7.7 &3.259  &A &1.14$\pm$0.15 & FIRST image & 0.9		 \\
J120319.1-011253 &114.3$\pm$7.4 &142.8$\pm$8.2 &110.2$\pm$8.6 &2.26$\pm$0.39 &C &0.82$\pm$0.15 &FIRST image &1.8	 \\
J125135.3+261457 &157.9$\pm$7.5 &202.3$\pm$8.2 &206.8$\pm$8.5 &3.675  &A &1.25$\pm$0.16 &FIRST cat &0.9 			 \\
J125759.5+224558 &272.4$\pm$7.3 &215$\pm$8.1 &137.8$\pm$8.7 &1.53$\pm$0.3 &B &0.85$\pm$0.14 &FIRST image &	0.9 \\
J131540.6+262322 &94.1$\pm$7.4 &116.1$\pm$8.2 &108.6$\pm$8.7 &2.4417 &C &0.96$\pm$0.15 &FIRST image & 1.2	 \\
J132427.0+284449 &342.4$\pm$7.3 &371$\pm$8.2 &250.9$\pm$8.5 &1.6760  &A &1.95$\pm$0.17 & FIRST cat &3.9 			 \\
J133255.7+342207 & 164.3$\pm$7.5 & 186.8$\pm$8.1 & 114.9$\pm$8.7 & 2.9268 & C & 1.00$\pm$0.16 &FIRST image & 1.9 \\
J133542.9+300401 & 136.6$\pm$7.2 & 145.7$\pm$8.0 & 125.0$\pm$8.5 & 2.685 &A & 0.14$\pm$0.02 & N17 obs   &            \\
J133649.9+291800 &294.1$\pm$6.7 &286$\pm$7.6 &194.1$\pm$8.2 &2.2024&A &0.87$\pm$0.15 &FIRST image &	1.2	 \\
J134429.4+303034 &462$\pm$7.4 &465.7$\pm$8.6 &343.3$\pm$8.7 &2.301&A&1.29$\pm$0.16 & FIRST cat &0.6			 \\
J142413.9+022303 &112.2$\pm$7.3 &182.2$\pm$8.2 &193.3$\pm$8.5 &4.243 &A &0.79$\pm$0.16 &FIRST image & 2.2		 \\
J142935.3-002836 &801.8$\pm$6.6 &438.5$\pm$7.5 &199.8$\pm$7.7 &1.027 &A &2.8$\pm$0.67 & M14 &		 \\
J230546.2-331038 &76.8$\pm$5.6 &110.9$\pm$5.9 &110.4$\pm$7 &3.67$\pm$0.56 &C &0.37$\pm$0.07 &ATCA image&	2.8	 \\
J232531.3-302235 &175.5$\pm$4.3 &227.0$\pm$4.7& 175.7$\pm$5.7 & 2.8$\pm$0.46 &C &0.105$\pm$0.03& ATCA image & 1.9 \\
J232900.6-321744 &118.3$\pm$4.7 &141.2$\pm$5.2 &119.7$\pm$6.4 &2.81$\pm$0.46 &C &0.31$\pm$0.06 &ATCA image	&	1.6 \\
\hline
\end{tabular}

\end{center}
N17: \cite{Nayyeri2017}, M14: \cite{Messias2014}; 
$^{\dag}$ spectroscopic redshifts of H-ATLASJ130333.1+244643, H-ATLASJ131540.6+262322 and H-ATLASJ133255.7+342207 are taken from \cite{Neri2020}.
\end{table*}
The FIRST (Faint Images of the Radio Sky at Twenty-cm; \citealt{Becker1995}) survey, performed with the VLA reaches a typical rms of 0.15 mJy beam $^{-1}$ and a resolution $\sim$ 5 arcsec comparable with those of our observations. FIRST overlaps
with the Equatorial and North Galactic Pole H-ATLAS fields, thus complementing our search for radio counterparts for the \cite{Negrello2017} sample. The cross-match with the FIRST catalogue within 10 arcsec of the \textit{Herschel}-ATLAS position yields 8 potential counterparts, while for 12 objects we find a clear $\gtrsim 3\,\sigma$ signal in the FIRST maps, within the H-ATLAS beamsize.
For one of the matched source, namely H-ATLAS J142935.3-002836, the FIRST flux density value has been replaced with the more reliable one coming from the JVLA 7 GHz high resolution ($\sim$ 0.3 arcsec) and high sensitivity (10 $\mu$Jy beam$^{-1}$) observations described in \cite{Messias2014} (converted at 1.4 GHz).

Observations at 6.89 GHz from the JVLA are available for one additional source (H-ATLASJ133542.9+300401) and are described in details by \cite{Nayyeri2017}. The image reaches an rms of 7.8 $\mu$Jy beam$^{-1}$ and a beam size of 1.01$\times$0.81 arcsec; for our analysis we use the flux density in the above paper already reported at 1.4 GHz (see Table 1 in \citealt{Nayyeri2017}).

H-ATLASJ090311.6+003907 (SDP.81) is detected in the FIRST catalogue, was observed by the Extended Very Large Array (EVLA) at 8.4 GHz (\citealt{Valtchanov2011}, \citealt{Rybak2015}). However, its radio emission is likely to be originated from the AGN hosted by the foreground lens (\citealt{Tamura2015a}, \citealt{Rybak2020}); for this reason it has been excluded a priori as a possible counterpart.

\subsection{Counterparts selection}
The detected sources display a variety of radio morphologies: some of them show a compact radio emission, others instead feature a more extended structure or even multiple components, but the angular resolution of the observations is not enough to resolve any of the arcs possibly associated to lensing effects.

For this reasons, in order to establish whether a radio source is correctly assigned as a counterpart of our selected Herschel sample, we selected radio detections with at least $> 3\sigma_{\rm image}$ confidence level and then consider their position with respect to the optical/IR and the mm high angular resolution imaging.

We consider as a counterpart only the component nearest to the SPIRE peak position, whose radio detected emission overlaps with sources detected the in NIR or mm, even though this can result in a underestimation of the radio luminosity in case of multiple component radio sources. In the NIR we exploit HST/WFC3 wide-J filter F110 maps described in \cite{Negrello2014} and more recent snapshot observations covering the Equatorial and Southern fields (PI: Marchetti L., 2019), both reaching an angular resolution of $\sim$ 0.13 arcsec. The remaining sources are observed by the UK Infrared Deep Sky Survey Large Area Survey (UKIDSS-LAS; \citealt{Lawrence2007}) and the VISTA Kilo-Degree Infared Galaxy Survey (VIKING; \citealt{Edge2013}), reaching angular resolutions $\lesssim 1$ arcsec.
Millimetric high resolution images come from the ALMA Science Archive: some of the sources located in the South Galactic Pole have been target of Band 4 follow-ups at $\sim$ 1.7 arcsec, while for objects in the Northern and Equatorial fields we make use of the images found in bands 4, 6 and 7, with highest angular resolutions spanning from $\sim1.2$ to $<0.1$ arcsec.
For sources without high resolution imaging we compared the uncertainties on the radio positional accuracy
(\citealt{Becker1995}) with respect to the one from SPIRE/250 $\mu$m (\citealt{Bourne2011}), with a $\sim 3\sigma$ tolerance. We include only objects showing a superimposition between the two centroids and/or between the SPIRE positional uncertainty and the radio detection. 

Following the above criteria we find 11 and 17 sources in ATCA and FIRST respectively, showed in Fig.~\ref{fig:matches} and Fig. ~\ref{fig:first_matches}. Please note that the HST/WFC3 maps reported here do not take into account the well known positional offset between ALMA and HST counterparts (\citealt{Dunlop2016}, \citealt{Rujopakarn2016}). 

Therefore, combining ATCA follow-up and FIRST cross-matches we collected radio measurements for 28 out of the 80 galaxies in the original sample by \cite{Negrello2017}, that will constitute our reference in the present paper.

For the remaining objects we define an upper limit for detection at 3$\sigma$ from the noise at the position of the sources in the FIRST images and as 3$\sigma_{\rm image}$ for ATCA images. Table \ref{tab:matches} summarises the radio and FIR photometry for our sample. 

Finally, few considerations need to be done. First, we would like to stress that in this paper we rely only on the available archived radio data without focusing into descriptions of the single objects. In our analysis we take into account the information available in the literature for the confirmed cases, but for the remaining unconfirmed ones in absence of an accurate lens modelling and/or higher angular resolution observations, it is not possible to confirm whether or in which fraction the radio emission is associated to the lensed object.
This is also valid for the SPIRE flux densities in the unconfirmed cases. Indeed, as showed in Figures \ref{fig:matches} and \ref{fig:first_matches}, multiple sources entering the \textit{Herschel} beam are detected in the ALMA (sub-)mm maps, leading to a possible overestimation of the effective FIR luminosity of the actual counterpart. A fraction of low resolution sub-mm detected sources is indeed expected to be composed by multiple ALMA sources (\citealt{Hodge2013}, \citealt{Smail2014}, \citealt{Bussmann2015}). Here we assume that the observed SPIRE flux density is mostly originated by the same object and that no strong FIR contamination from possible nearby sources is present. 
In any case, this should not affect the objects already confirmed to be lensed. In fact, we expect the FIR luminosity of the lensed galaxies in our sample to be magnified by a factor spanning the range $\sim 3-10$ (\citealt{Negrello2017}, \citealt{Enia2018}), with a negligible contribution from possible additional unlensed sources entering the \textit{Herschel} beam.

\subsection{Physical properties of the sample}\label{subsec:properties}

The rest-frame radio luminosity $L_{\rm 1.4 GHz}$ at 1.4 GHz for each source (see Table \ref{tab:properties}) is computed as
\begin{equation}
    L_{\nu,e}= \frac{4 \pi D_{L}^2(z)}{(1+z)^{1+\alpha}} \left( \frac{\nu_e}{\nu_o} \right)^{\alpha} S_{\nu,o},
\end{equation}
where $S_{\nu} \propto \nu ^{\alpha}$ is the monochromatic flux density at a certain frequency and $\alpha$ is assumed to be -0.7 as the typical value at 1.4 GHz for FIRST (\citealt{Kimball2008}), $\nu_e$ and $\nu_o$ are the emitted and the observed frequency, and $D_L$ is the luminosity distance computed for each redshift according to the adopted $\Lambda$CDM cosmology.
Our 28 objects span a range $1.9\times10^{24}\lesssim L_{\rm 1.4 GHz}\lesssim 1.8\times 10^{26}$ W Hz$^{-1}$ (uncorrected for the lens magnification factor $\mu$), with a median value of $L_{\rm 1.4 GHz}\sim 2.4\times 10^{25}$ W Hz$^{-1}$.

\begin{figure}
    \centering
    \includegraphics[width=0.47\textwidth]{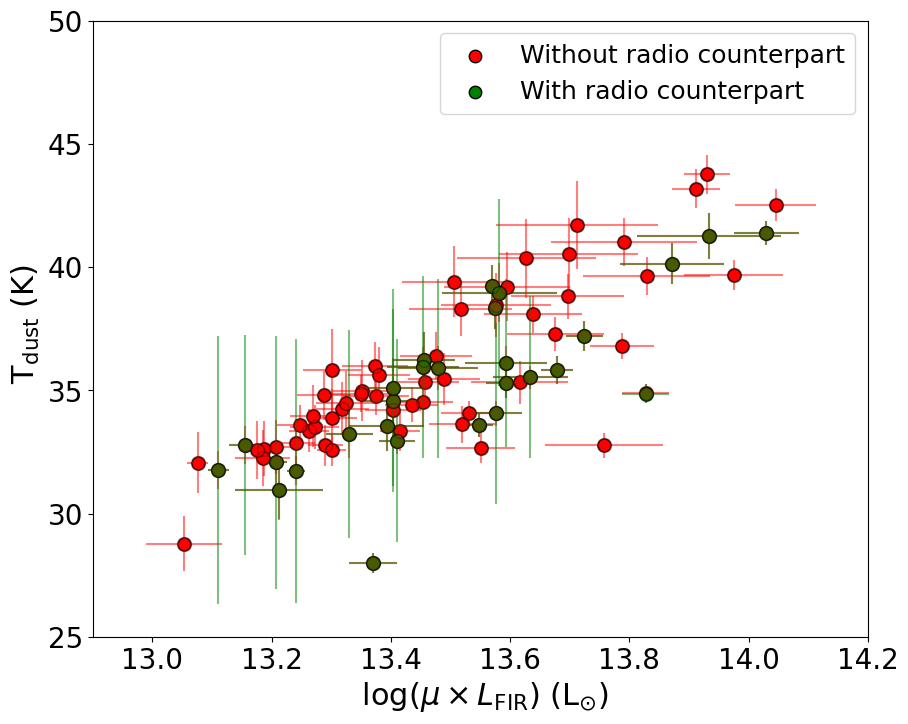}
    \caption{Dust temperatures versus FIR luminosities (uncorrected for magnification) for the 80 (candidate) strongly lensed galaxies from \protect\cite{Negrello2017}. Green and red dots show respectively sources with and without radio counterpart.}
    \label{fig:T_dust}
\end{figure}

The FIR luminosity $L_{\rm FIR}$ (see Table \ref{tab:properties}) is computed for each source in the main sample of \cite{Negrello2017} by fitting the \textit{Herschel}/SPIRE photometry described in Section \ref{subsec:h_atlas_sample}. We use a single-temperature modified black body under the optically-thin approximation with dust emissivity index $\beta =$ 1.5 (\citealt{Nayyeri2016}, \citealt{Negrello2017}), the spectrum normalisation and the dust temperature (T$_{\rm dust}$) are kept as free parameters. The model ($S_{\nu, best}$) which minimises the $\chi^2$ is then integrated over the wavelength range 8-1000 $\mu$m as follows: 
\begin{equation}
    L_{\rm FIR}= \frac{4\pi D_L^2}{(1+z)} \int^{1000\mu \rm m}_{8 \mu \rm m} S_{\nu, best} d\nu.
\end{equation}
The resulting FIR luminosities (uncorrected for lens magnification) and dust temperatures are showed in Figure \ref{fig:T_dust}. The FIR luminosities span the range $1.3\times 10^{13}\lesssim L_{\rm FIR} \lesssim 1.1\times 10^{14}$ $L_{\odot}$, with a median value of $L_{\rm FIR} \sim 3.5 \times 10^{13}$ $L_{\odot}$. The median value of the dust temperature for the total sample of  candidate strongly lensed galaxies is $T_{\rm dust}= 35.2 \pm 2.1$, consistently with what was found by \cite{Negrello2017} for the same sample.

\begin{table}
\centering
\caption{$1.4$ GHz and FIR luminosities (\textit{uncorrected} for lensing magnification effects) and q$_{\rm FIR}$ parameter for the 28 galaxies in our sample.}
\label{tab:properties}

\begin{tabular}{lcccc}
\hline
$\#$H-ATLAS ID   & $\log(\rm L_{\rm 1.4GHz})$ & $\log(\rm L_{\rm FIR})$ & q$_{\rm FIR}$\\
 & [$\rm L_{\odot}$]& [W Hz$^{-1}$] & \\
 \hline
J000007.4-334059   &  25.86$\pm$0.04	   &  13.40$\pm$0.05	   & 1.55$\pm$0.15  \\
J000722.1-352014   &  24.62$\pm$0.06	   &  13.11$\pm$0.02	   & 2.49$\pm$0.13  \\
J000912.7-300807   &  24.29$\pm$0.10	   &  13.24$\pm$0.01	   & 2.96$\pm$0.23  \\
J005132.8-301848   &  25.13$\pm$0.08	   &  13.59$\pm$0.07	   & 2.47$\pm$0.24  \\
J010250.8-311723   &  24.77$\pm$0.07	   &  13.41$\pm$0.03	   & 2.64$\pm$0.17  \\
J012407.3-281434   &  25.37$\pm$0.05	   &  13.58$\pm$0.04	   & 2.20$\pm$0.14  \\
J013004.0-305513   &  25.07$\pm$0.05	   &  13.15$\pm$0.03	   & 2.08$\pm$0.12  \\
J013239.9-330906   &  24.98$\pm$0.10	   &  13.48$\pm$0.07	   & 2.50$\pm$0.28  \\
J085358.9+015537   &  25.46$\pm$0.06	   &  13.68$\pm$0.03	   & 2.22$\pm$0.14  \\
J090302.9-014127   &  25.09$\pm$0.15	   &  13.72$\pm$0.03	   & 2.64$\pm$0.35  \\
J090740.0-004200   &  25.19$\pm$0.05	   &  13.57$\pm$0.01	   & 2.39$\pm$0.12  \\
J091043.0-000322   &  25.22$\pm$0.08	   &  13.55$\pm$0.02	   & 2.33$\pm$0.20  \\
J091840.8+023048   &  26.25$\pm$0.02	   &  13.39$\pm$0.06	   & 1.14$\pm$0.15  \\
J114637.9-001132   &  25.86$\pm$0.06	   &  14.03$\pm$0.05	   & 2.17$\pm$0.18  \\
J120319.1-011253   &  25.54$\pm$0.08	   &  13.40$\pm$0.06	   & 1.86$\pm$0.23  \\
J125135.3+261457   &  26.01$\pm$0.05	   &  13.87$\pm$0.09	   & 1.86$\pm$0.23  \\
J125759.5+224558   &  25.01$\pm$0.07	   &  13.21$\pm$0.02	   & 2.20$\pm$0.17  \\
J131540.6+262322   &  25.51$\pm$0.07	   &  13.21$\pm$0.07	   & 1.70$\pm$0.23  \\
J132427.0+284449   &  25.46$\pm$0.04	   &  13.37$\pm$0.04	   & 1.91$\pm$0.12  \\
J133255.7+342207   &  25.42$\pm$0.07	   &  13.33$\pm$0.04	   & 1.92$\pm$0.18  \\
J133542.9+300401   &  24.75$\pm$0.07	   &  13.45$\pm$0.05	   & 2.70$\pm$0.20  \\
J133649.9+291800   &  25.37$\pm$0.07	   &  13.59$\pm$0.03	   & 2.22$\pm$0.18  \\
J134429.4+303034   &  25.59$\pm$0.05	   &  13.83$\pm$0.04	   & 2.24$\pm$0.15  \\
J142413.9+022303   &  25.95$\pm$0.09	   &  13.93$\pm$0.12	   & 1.99$\pm$0.34  \\
J142935.3-002836   &  25.13$\pm$0.10	   &  13.57$\pm$0.01	   & 2.44$\pm$0.23  \\
J230546.2-331038   &  25.49$\pm$0.06	   &  13.58$\pm$0.1	   & 2.10$\pm$0.27  \\
J232531.3-302235   &  24.69$\pm$0.11	   &  13.63$\pm$0.06	   & 2.95$\pm$0.30  \\
J232900.6-321744   &  25.15$\pm$0.06	   &  13.45$\pm$0.06	   & 2.30$\pm$0.19  \\

\hline
\end{tabular}
\end{table}

\section{The FIR-radio correlation for (candidate) lensed galaxies}\label{sec:FIRRC}

In this section we explore the correlation between the radio and the FIR luminosities for our sample of 28 (candidate) lensed dusty star-forming galaxies, focusing on the observed quantities uncorrected for lensing magnification effects. For each galaxy we computed the $q_{\rm FIR}$ parameter following equation \ref{eq:q_ir}.

\begin{figure*}
	\includegraphics[width=12cm]{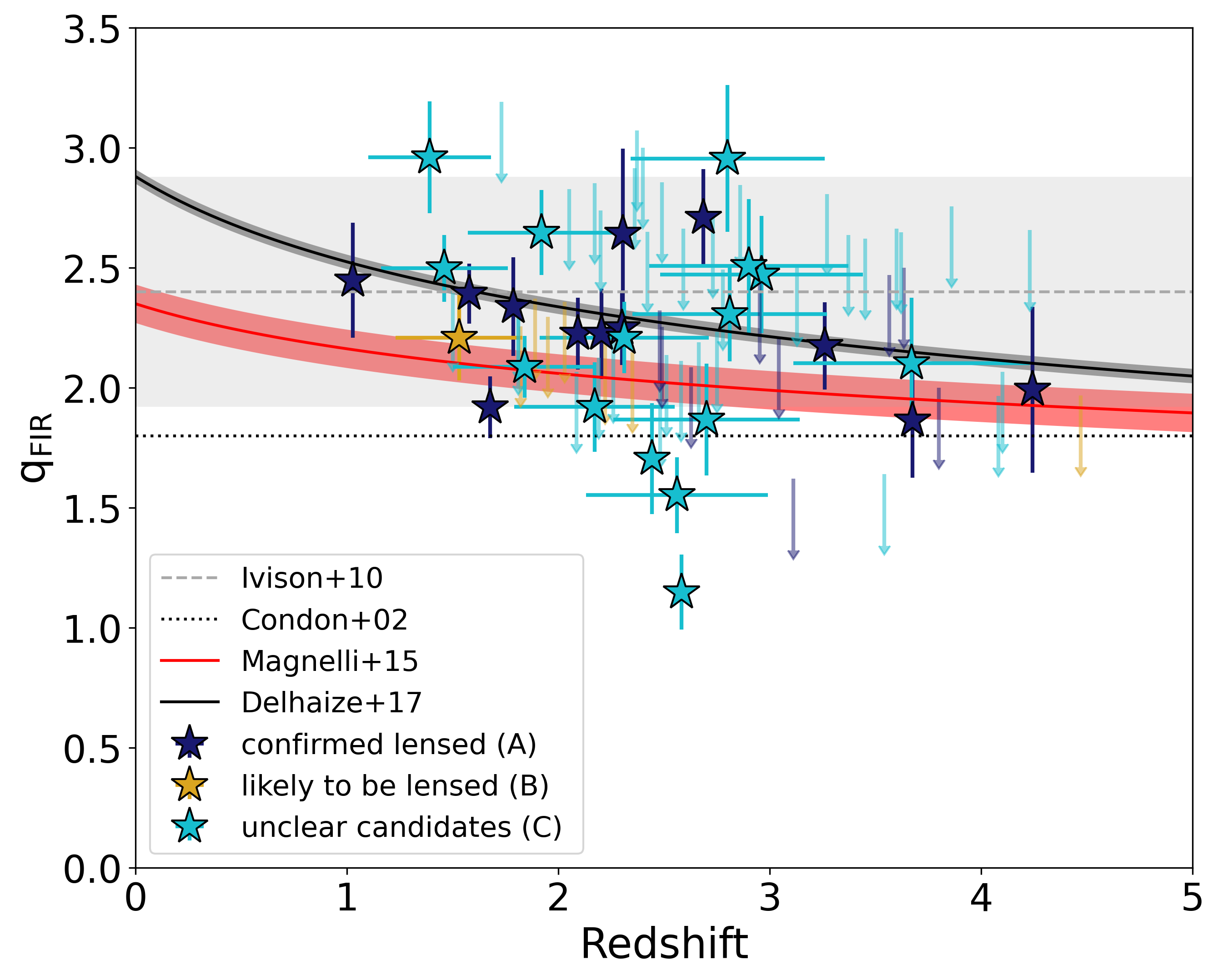}    
 \caption{ $q_{\rm FIR}$ parameter (see equation \ref{eq:q_ir}) as a function of redshift. The grey dashed line corresponds to the median $q_{\rm FIR}$ for star-forming galaxies as defined by \protect\cite{Ivison2010b}, the grey shaded area represents the $2\sigma$ dispersion.
 Red and black shaded areas represent respectively the relation by \protect\cite{Magnelli2015} and \protect\cite{Delhaize2017}. Stars are the $28$ galaxies in our sample: (blue) the confirmed lensed objects, (yellow) the likely lensed objects, and (cyan) the uncertain objects. Arrows show the 3$\sigma$ upper limits for the remaining undetected sources.
    }
    \label{fig:firrc_z}
\end{figure*}

\begin{figure*}
    \centering
    \includegraphics[width=12cm]{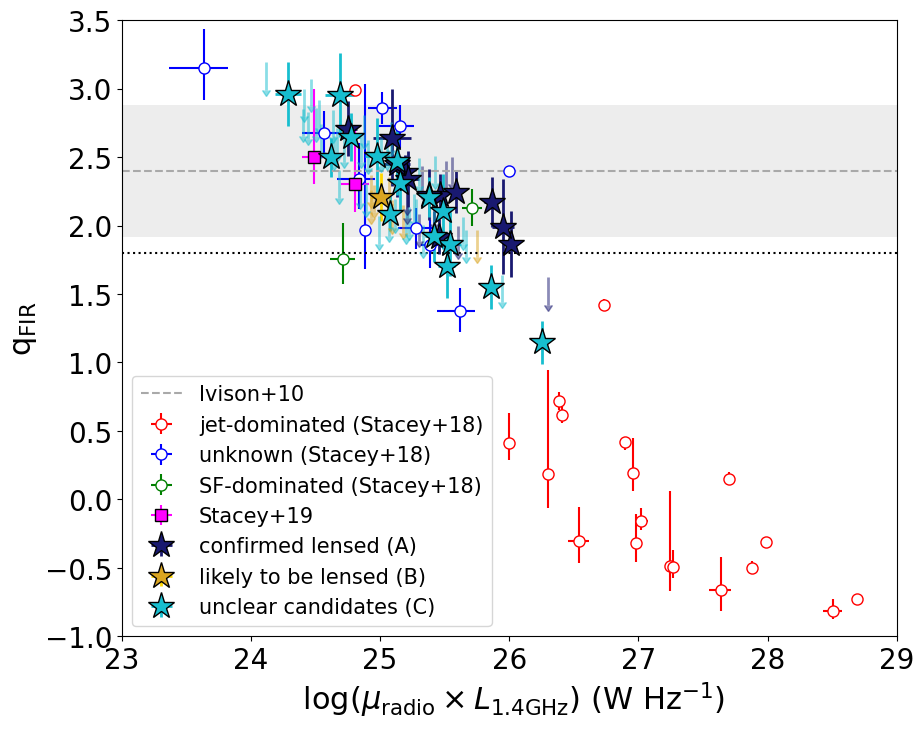}
    \caption{The $q_{\rm FIR}$ parameter, computed using the FIR luminosity, as a function of the logarithm of the 1.4 GHz luminosity (not corrected for lens magnification). Circles are the \protect\cite{Stacey2018} lensed quasars classified according to the origin of their radio emission: (red) jets, (green) star formation, and (blue) unknown. Squares are from \protect\cite{Stacey2019}. Arrows show the 3$\sigma$ upper limits for the remaining undetected sources.}
    \label{fig:firrc_Lradio}
\end{figure*}

Figure \ref{fig:firrc_z} shows the $q_{\rm FIR}$ parameter as a function of redshift. We find a weak yet appreciable decline of $q_{\rm FIR}$ with increasing $z$. Our result is compared with the redshift evolution reported by \cite{Magnelli2015} and \cite{Delhaize2017} (red and green shaded areas respectively) and with the median value of the q$_{\rm FIR}$ parameter for star forming radio galaxies from \cite{Ivison2010b}. 
\cite{Magnelli2015} studied a mass-selected sample of star-forming galaxies up to $z\sim 2$, finding a slight evolution $q_{\rm FIR} (z)= (2.35 \pm 0.08)(1 + z)^{-0.12\pm0.04} $; in Fig.\ref{fig:firrc_z} their relation is extended up to higher redshifts to ease the comparison with our data. 
Similar results were obtained by \cite{Delhaize2017} from a radio-selected sample of star-forming galaxies extending up to $z\sim 6$, as described by the relation $q_{\rm FIR} (z)=(2.52 \pm 0.03)(1 + z)^{-0.21\pm0.01}$.
The majority of our sources show values of $q_{\rm FIR}$ within the 2$\sigma$ interval around the median value of \cite{Ivison2010b} for star-forming galaxies. Three sources lie below the limit of $q_{\rm FIR}\sim 1.8$ established by \citealt{Condon2002} to separate between sources with radio emission powered by star formation and by AGN, respectively. The AGN-powered objects are all located at redshifts $z\gtrsim 2$.

Figure \ref{fig:firrc_Lradio} shows the dependence of the $q_{\rm FIR}$ parameter on the radio luminosity (uncorrected for lensing magnification). We find a clear tendency of the $q_{\rm FIR}$ parameter to decrease with increasing radio power. The radio excess follows from the definition given in Eq. \ref{eq:q_ir} and, consequently, this observed trend can be ascribed to the presence of an AGN. 
Such a behaviour is similar to what have been revealed by \cite{Stacey2018} observing a sample of strong gravitationally-lensed quasars. Their sample includes 104 quasars lensed by foreground galaxies, listed in the SQLS catalogue and CASTLES database (\citealt{Munoz1999}, \citealt{Kochanek1999}, \citealt{Inada2012}) and detected in a variety of optical and radio surveys.
For clarity we report in the Figure their 31 detected sources that could be divided in three categories: jetted quasars, where high-resolution radio data confirmed the emission to be associated with AGN jets; non-jetted quasars, dominated by radio emission triggered by star-formation; quasars where the origin of the radio emission is unknown. They assumed a magnification factor of $\mu = 10^{+10}_{-5}$ for the majority of their sample, and the estimated median value of the total infrared luminosity amounts to $3.6^{+4.8}_{-2.4}x10^{11}\,L_{\odot}$. Two additional sources are the gravitationally lensed radio-quiet quasars targeted by the LOFAR Two-metre Sky Survey (LoTSS) and detected by \textit{Herschel} described in \cite{Stacey2019}.

19 out of the 31 detected lensed quasars from \cite{Stacey2018} are located below the threshold of 1.8 and are mostly (with only one exception) classified as jet-dominated. For the remaining sources, only two quasars are confirmed to be star formation-dominated sources and are found to be above the 1.8 threshold within their uncertainties. The nature of the radio emission of the remaining one third of the sample (10 sources) is still uncertain.
Only the $25\%$ of their sample has $q_{\rm FIR}<1.8$ and lie in the same region of the jet-dominated quasars of \cite{Stacey2018}, actually aligning onto their trend of $q_{\rm FIR}$ with radio luminosity. Most relevantly, the majority of our sources lie ($\sim 90 \%$) in the region where radio emission is dominated by star formation, thus complementing the radio/optical selection adopted in \cite{Stacey2018} and actually extending to lower radio luminosities a consistent $q_{\rm FIR}$ vs. radio luminosity relation.

Finally, our assumption that the FIR luminosity is not strongly affected by nearby objects entering the \textit{Herschel} beam is reasonable. If not, the trend in Figure \ref{fig:firrc_Lradio} should show a clear offset with respect to the confirmed lensed galaxies and/or with the points from \cite{Stacey2018,Stacey2019}. Another support to this assumption is given by the fact that our points are consistent within the uncertainties with the span of the $q_{\rm FIR}- L_{1.4 \rm GHz}$ relation.

\section{Physical interpretation}\label{sec:evolutionary_scenario}

\begin{figure*}
    \centering
    \includegraphics[width=12cm]{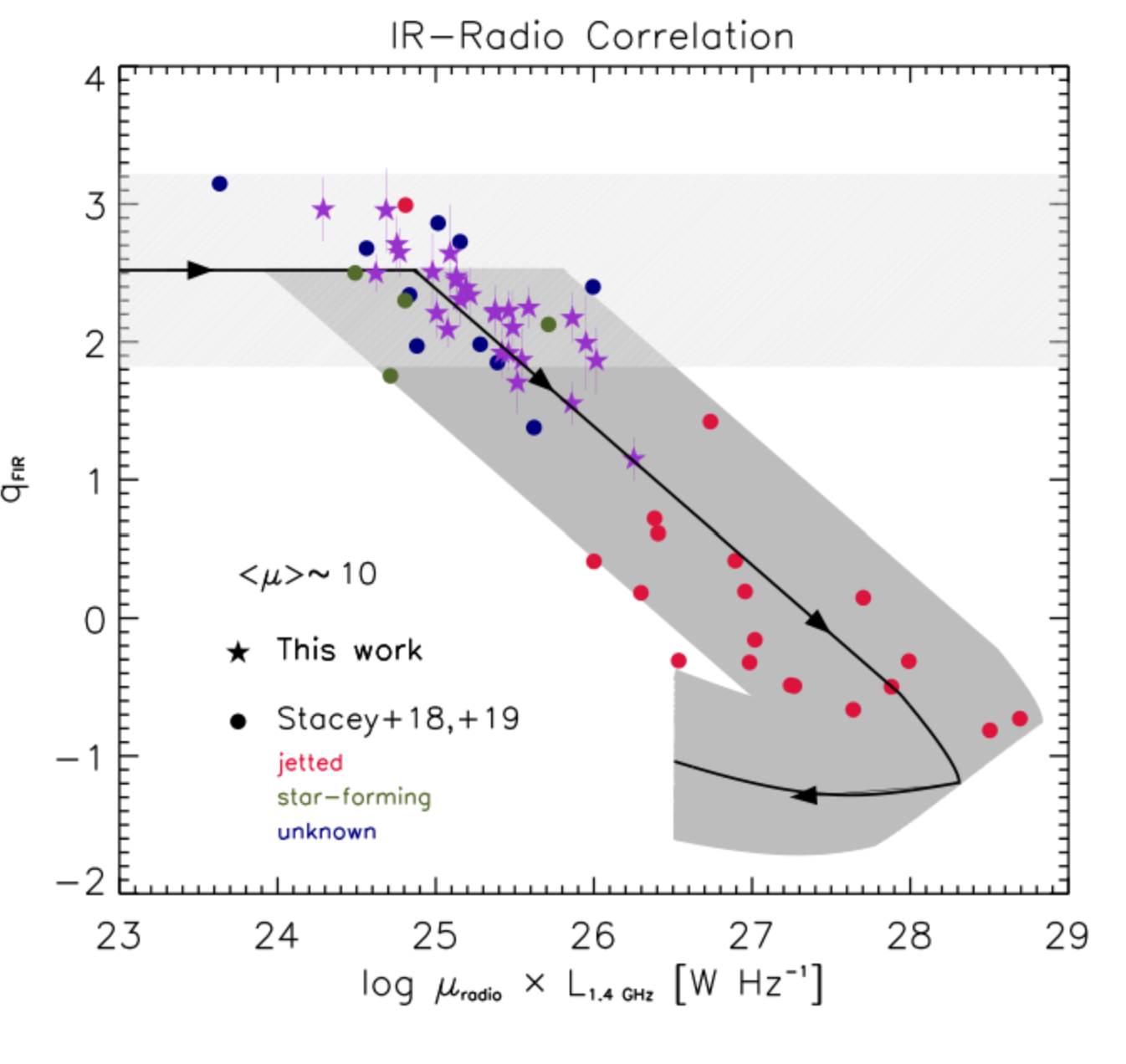}
    \caption{The FIR-radio correlation in terms of the $q_{\rm FIR}$ vs. radio luminosity (uncorrected for lensing magnification). Data from this work (stars) and from Stacey et al. (\citeyear{Stacey2018}, \citeyear{Stacey2019}. Circles: red for jetted source, green for star-forming galaxies and blue for uncertain ones) are compared with the prediction of the in-situ galaxy formation scenario by \protect\cite{Lapi2014, Lapi2018} for radio loud AGNs. Specifically, the dark grey area represent the locus of radio-loud AGNs and the light shaded area that for star-forming galaxies without a significant AGN contribution to the radio emission. The black line with arrows illustrates the typical evolution of an individual radio-loud AGN at $z\sim 2$. An average magnification $\langle\mu\rangle\sim 10$ (as estimated by \protect\citealt{Stacey2018}) has been applied to the model predictions for fair comparison with the data of gravitationally lensed sources.}
    \label{fig:firrc_model}
\end{figure*}

In the way of providing a physical interpretation, in Fig. \ref{fig:firrc_model} we compare the observed FIR-radio correlation from this work and from \cite{Stacey2018,Stacey2019} to the in-situ galaxy formation scenario by \cite{Lapi2014,Lapi2018} (see also \citealt{Mancuso2017}, \citealt{Pantoni2019}). This scenario envisages star formation and (super)massive black hole (BH) accretion in galaxies to be essentially in-situ and time-coordinated processes, triggered by the fast collapse of baryons in the host dark matter halos and subsequently controlled by self-regulated baryonic physics, in particular by energy/momentum feedback from SNe and from the central active nucleus.

The evolution of an individual massive galaxy (say the high-redshift star-forming progenitor of a present-day elliptical) predicted by the model consists of different stages (see \citealt{Lapi2018}, their Fig. 1 for a schematic illustration). Early on, the balance between cooling, infall, compaction and stellar feedback processes sets in a strong and dust-enshrouded star formation activity (SFR$\sim$ several $10^2$ M$_\odot$ yr$^{-1}$), with a roughly constant behaviour in time (stellar mass increases almost linearly). Meanwhile, in the inner, gas-rich galaxy regions the central BH undergoes an exponential growth in Eddington-limited conditions. In this stage the system behaves as a bright IR/(sub)mm galaxy with an X-ray active nucleus. When the central BH has grown to a significant mass, its energy output becomes so large as to eject gas and dust from the host and eventually quench the star formation and reduce the accretion onto itself to sub-Eddington values. In this stage the system behaves as a powerful AGN with possible residual star formation. Thereafter, the stellar population evolves almost passively, and the system behaves as a red and dead massive quiescent galaxy; its further evolution in mass and size toward the present is mainly due to minor dry merger events.

In the early stages of the evolution when the central BH is still small and the nuclear power quite limited, the radio emission is mostly associated to the star formation in the host, implying modest $L_{1.4\,\rm GHz}$ and standard values $q_{\rm FIR}\approx 2.5$. Later on, when the BH mass has increased to substantial values, the radio emission from the nucleus progressively overwhelms that from the star formation, driving $q_{\rm FIR}$ toward values appreciably smaller than $2.5$. Values smaller even than $1.8$ are obtained when jetted emission is produced, preferentially in the late-stage of the evolution: this is because extraction of rotational energy to drive jets is favoured in sub-Eddington conditions that sets in when the gaseous environment around the BH has been partially cleaned by the feedback from the nucleus itself (see above).

The typical model evolutionary track of a radio-loud AGN in the $q_{\rm FIR}$ vs. radio luminosity diagram is illustrated by the black line with arrows in Fig. \ref{fig:firrc_model}; the dark shaded area is the locus expected for such objects taking into account the relative contribution of sources with different masses, weighted by their statistics and timescales. For reference, the light shaded area refers to the typical $q_{\rm FIR}$ parameter of a star-forming galaxies with negligible contribution from AGNs to the radio emission. The agreement of the model prediction with the data is pleasingly good, and testifies that the behaviour of the FIR-radio correlation is physically driven by the onset of jetted radio emission, and consistent with a co-evolution scenario between star formation and BH growth. A further, more detailed comparison between such a scenario and data will be pursued in a forthcoming paper, where we will estimate physical quantities (e.g., SFR, stellar mass, stellar ages, etc.) for the galaxy sample presented in this work via broadband spectral energy distribution fitting.

\section{Conclusions}\label{sec:conclusions}

Our analysis exploited the FIR-selected, \textit{Herschel}-ATLAS (H-ATLAS) candidate strongly-lensed galaxy sample by \cite{Negrello2017} to investigate the FIR-radio correlation out to redshift $z\sim 4$. Specifically:

\begin{itemize}

\item We cross-matched the \cite{Negrello2017} sample with the FIRST survey at $1.4$ GHz and run dedicated follow-up with ATCA at $2.1$ GHz, finding $11$ and $16$ matches, respectively; the addition of another source observed at $\sim 7$ GHz with the JVLA (analysed in detail by \citealt{Nayyeri2017}) leaded us to a sample of $28$ candidate lensed dusty star-forming galaxies with a radio counterpart over the redshift range $1\lesssim z \lesssim 4$.

\item We derived the radio and the integrated FIR luminosities for the selected sample (uncorrected for lensing magnification), that feature radio luminosities in the range $1.9\times10^{24}\lesssim L_{\rm 1.4\,GHz}$ [W Hz$^{-1}$] $\lesssim 1.8\times 10^{26}$ and integrated FIR luminosity in the range $1.3\times 10^{13}\lesssim L_{\rm FIR}/L_{\odot}\lesssim 1.1\times 10^{14}$.

\item By taking advantage of the source brightness possibly enhanced by lensing magnification we identified a weak evolution with redshift out to $z\lesssim 4$ of the FIR-to-radio luminosity ratio $q_{\rm FIR}$, consistent with previous determinations at lower redshift based on different selections.

\item We found that the $q_{\rm FIR}$ parameter as a function of the radio power $L_{1.4\,\rm GHz}$ displays a clear decreasing trend, similarly to the lensed quasars selected in optical/radio by \cite{Stacey2018}, yet covering a complementary region in the $q_{\rm FIR}-L_{1.4\,\rm GHz}$ diagram. 

\item We interpreted the behavior of the FIR-radio correlation according to an in-situ galaxy evolution scenario, as the result of the transition from an early dust-obscured star-forming phase (mainly pinpointed by our FIR selection) to a late radio-loud quasar phase (preferentially sampled by the \citealt{Stacey2018} selection).  

\end{itemize}

\section*{Acknowledgements}
This paper makes use of the following ALMA data: 2013.1.00164.S, 2013.1.00358.S, 2015.1.00415.S, 2015.1.01455.S, 2016.1.00282.S, 2016.1.00450.S, 2017.1.00027.S, 2018.1.00526.S and 2019.1.01477.S. ALMA is a partnership of ESO (representing its member states), NSF (USA) and NINS (Japan), together with NRC (Canada), MOST and ASIAA (Taiwan), and KASI (Republic of Korea), in cooperation with the Republic of Chile. The Joint ALMA Observatory is operated by ESO, AUI/NRAO and NAOJ.
We acknowledge financial support from the grant PRIN MIUR
2017 prot. 20173ML3WW 001 and 002 ‘Opening the ALMA window on the cosmic evolution of gas, stars, and supermassive black holes’. AL is supported by the EU H2020-MSCAITN-2019 project 860744 ‘BiD4BEST: 
Big Data applications for Black hole Evolution STudies’.
We acknowledge the referee, J. P. McKean, for the useful comments.

\section*{Data Availability}

The data underlying this article are the following:
\begin{itemize}
    \item The public catalogue of 80 candidate gravitationally lensed galaxies extracted from the full H-ATLAS survey, published by \cite{Negrello2017};
    \item The \textit{Herschel}/SPIRE images at 250$\mu$m, produced from the H-ATLAS DR2 maps (\url{https://www.h-atlas.org/public-data/download});
    \item The Australian Telescope Compact Array data products from the Australia Telescope Online Archive, publicly available at the following link: \url{https://atoa.atnf.csiro.au/query.jsp}, project code: C3215;
    \item The last version of the Very Large Array FIRST survey catalogue (2014 Dec 14, \citealt{Helfand2015}), accessible through VizieR: \url{https://vizier.u-strasbg.fr/viz-bin/VizieR-3};
    \item The Very Large Array FIRST survey maps extracted from the cutout service available at the following link: \url{https://third.ucllnl.org/cgi-bin/firstcutout}.
    \item HST images taken from MAST: Barbara A. Mikulski Archive for Space Telescopes, \url{https://mast.stsci.edu/}.
    \item VIKING and UKIDSS-LAS cutouts, generated from \url{https://alasky.u-strasbg.fr/hipsimage-services/hips2fits}.
    \item ALMA images available in the ALMA Science Archive, project codes for the adopted images are: 2013.1.00164.S, 2013.1.00358.S, 2015.1.00415.S, 2015.1.01455.S, 2016.1.00282.S, 2016.1.00450.S, 2017.1.00027.S, 2018.1.00526.S and 2019.1.01477.S.
\end{itemize}

\bibliographystyle{mnras}
\bibliography{Giulietti21}

\bsp	
\label{lastpage}
\end{document}